\documentclass[twocolumn]{aastex63}
\usepackage{lineno}

\newcommand\msunyr{$M$\mbox{$_{\normalsize\odot}$} \rm{yr}$^{-1}$}
\newcommand\msun{$M$\mbox{$_{\normalsize\odot}$}}

\newcommand\mini{$M_{\rm init}$}

\newcommand\mdot{$\dot{M}$}

\newcommand\lbol{$L$\mbox{$_{\rm bol}$}}

\newcommand\teff{$T_{\rm eff}$}

\newcommand\logl{$\log (L/L_\odot)$}

\newcommand\N[1]{{\color{blue} \bf #1}} %N8

\usepackage{soul}

\received{}
\revised{}
\accepted{\today}
\submitjournal{ApJ}

\shorttitle{High \mdot\ RSGs}
\shortauthors{Beasor \& Smith}
\graphicspath{{./}{figures/}}
\begin{document}

\title{ The extreme scarcity of dust-enshrouded red supergiants: consequences for producing stripped stars via winds}

\correspondingauthor{Emma Beasor}
\email{embeasor@gmail.com}

\author[0000-0003-4666-4606]{Emma R. Beasor}\altaffiliation{Hubble Fellow}
\affiliation{NSF's NOIRLab
950 N. Cherry Ave., Tucson, AZ 85719, USA}

\author[0000-0001-5510-2424]{Nathan Smith}
\affiliation{Steward Observatory, University of Arizona
933 N. Cherry Ave., Tucson, AZ 85721, USA}



\begin{abstract}

Quiescent mass-loss during the red supergiant (RSG) phase has been shown to be far lower than prescriptions typically employed in single-star evolutionary models. Importantly, RSG winds are too weak to drive the production of Wolf-Rayets (WRs) and stripped-envelope supernovae (SE-SNe) at initial masses of roughly 20--40$M_{\odot}$. If single-stars are to make WRs and SE-SNe, this shifts the burden of mass-loss to rare dust-enshrouded RSGs (DE-RSGs), objects claimed to represent a short-lived high mass-loss phase. Here, we take a fresh look at the purported DE-RSGs. By modeling the mid-IR excesses of the full sample of RSGs in the LMC, we find that only one RSG has both a high mass-loss rate (\mdot $\ge$ 10$^{-4}$ $M_{\odot}$ yr$^{-1}$) and a high optical circumstellar dust extinction (7.92 mag). This one RSG is WOH G64, and it is the only one of the 14 originally proposed DE-RSGs that is actually dust enshrouded. The rest appear to be either normal RSGs without strong infrared-excess, or lower-mass asymptotic giant branch (AGB) stars. Only one additional object in the full catalog of RSGs (not previously identified as a DE-RSG) shows strong mid-IR excess. We conclude that if DE-RSGs do represent a pre-SN phase of enhanced \mdot\ in single-stars, it is extremely short-lived, only capable of removing $\leq$2\msun\ of material. This rules out the single-star post-RSG pathway for the production of WRs, LBVs, and SE-SN. Single-star models should not employ \mdot-prescriptions based on these extreme objects for any significant fraction of the RSG phase.

\end{abstract}

%%%%%%%%%%%%%%%%%%%%%%%%%%%%%%%%%%%%%%%%%%%%%%%%%%

%%%%%%%%%%%%%%%%% BODY OF PAPER %%%%%%%%%%%%%%%%%%

\section{Introduction}
Under the single-star evolutionary paradigm, the red supergiant (RSG) phase is the expected final evolutionary point for stars with initial masses (\mini) between 8--25\msun\ \citep[e.g.][]{meynet2003stellar,heger2003how}, or even higher depending on the mass-loss rate employed. Eventually, they exhaust the fuel in their cores and end their lives as core-collapse supernovae (CCSNe). There are many direct detections confirming RSGs as progenitors of normal Type II-P supernovae (SNe II-P) \citep[e.g.][]{maund2005hubble,smartt2009death}. The precise type of SN produced depends strongly on the amount of hydrogen left in the envelope at the time of explosion, which is determined by the star's mass-loss rate (\mdot) history \citep[see][]{smith2014mass}.  SNe II-P may have a variety of ejected H envelope masses at the time of explosion, and if the H envelope is mostly or entirely  lost, progenitors can yield SNe of Types II-L or IIn, or even stripped-envelope  supernovae (SE-SNe) of Types IIb, Ib, and Ic.  For initial  masses of 8--25\msun, stars that have their H envelopes stripped in a binary system are likely progenitors of many stripped-envelope SNe, but the fraction of SE-SNe made by single stars (if any) depends most critically on the uncertain wind mass loss during the RSG phase.

Evolutionary models are able to make predictions about the numbers of stars expected to retain their H-envelope prior to SN \citep[e.g.][]{ekstrom2012grids,choi2016mist}.  However, these predictions strongly depend on  mass-loss rates and the way in which they are implemented. For stars in the mass range 8--30\msun, OB star winds on the main sequence are weak \citep{smith2014mass}. As such, it is mass-loss during the RSG phase which has the potential to most drastically alter their evolution. The precise mechanism driving RSG mass-loss is not fully understood.  It is likely a combination of radiation pressure on dust grains \citep{gehrz1971mass,wilson2000high} and pulsations \citep{yoon2010evolution} that lift mass to radii where dust can form, but so far no model can accurately match observations, and there is still no theoretically motivated  prescription for how RSG mass-loss rates depend on mass, luminosity, temperature, or pulsational instability.

Due to this uncertainty, stellar evolution models rely on empirical recipes to implement mass-loss and calculate evolution. There is now growing evidence that the most commonly used RSG \mdot-prescription, that of \citet{de1988mass}, overestimates the amount of mass lost for high-mass progenitors \citep{beasor2016evolution,beasor2018evolution,beasor2020new}. \citet{beasor2020new} provide a revised mass-loss rate prescription for RSGs that can be implemented in modern stellar evolution codes.   In these works, \citeauthor{beasor2020new} targeted 34 RSGs in 4 clusters (compared to a total of 10 field RSGs in the de Jager et al. sample), where the distance and age are known, and derived \mdot\ and luminosity values for each star. It was found that even at the highest luminosities, the quiescent RSG mass-loss rates are relatively low ($<$10$^{-5}$ \msun\ yr$^{-1}$), and would not be able to remove the star's H-envelope prior to explosion\footnote{Recent work by \citet{humphreys2020mass} also investigated RSG mass-loss using a similar sample of stars to that of \citet{beasor2020new}. Following  \citet{de1988mass}, those authors do not take into account the different initial masses for stars of a given luminosity, and instead derive an \mdot-prescription based only on luminosity. Although the range of derived mass-loss rates found by \citet{humphreys2020mass} agreed well with \citet{beasor2020new}, the mass-loss rates had a much larger dispersion above and below the Humphreys et al. prescription. \citet{beasor2020new} demonstrated that such a large dispersion arises if one neglects the fact that there is a range of different initial masses at the same luminosity, and that these stars of different masses actually have different mass-loss rates. \citet{beasor2020new} further showed that using a prescription that neglects this range of stellar masses at the same luminosity artificially inflates the mass lost during the RSG phase. }
%While the main conclusions of \citet{humphreys2020mass} were that the \citet{de1988mass} does not overestimate mass-loss, we note that the authors do not take into account the different initial masses for the stars within each cluster, and instead derive an \mdot-prescription that is based only on luminosity. As shown in Fig. 12 within \citeauthor{humphreys2020mass}, the trend of increasing offset for the \mdot-\lbol\ relation with initial mass was also observed. Thus, if the authors also derived instead an initial-mass dependent \mdot-prescription \citep[as in][]{beasor2020new} it is likely they would have also found the \citet{de1988mass} prescription overestimates the total mass lost during the RSG phase. }}. 

If single RSGs are to shed their H envelopes to make Wolf-Rayet (WR) stars and stripped-envelope SNe, one would need to invoke episodes with much stronger RSG winds.  Some models do adopt enhanced mass-loss rates that can remove some or all of the H envelope, motivated by a mass-loss rate prescription \citep{van2005empirical} derived from a subset of RSGs that have been interpreted as ``dust enshrouded RSGs" \citep[DE-RSGs hereafter, see also][]{sargent2010mass,sargent2011mass,rieber2012,blum2014,groenewegen2018luminosities}.  These stars are sometimes also described as being in a temporary ``superwind" phase.  Using the DE-RSG mass-loss prescription in models for single stars implicitly assumes that the DE-RSG phase is an episode that all single stars pass through.%; it remains possible, however, that the DE-RSGs might instead result from binary interaction.

The mass-loss rates for DE-RSGs \citep{van2005empirical,goldman2017wind} are claimed to be more than a factor of 10 higher than those seen among populations of normal RSGs in clusters \citep{beasor2020new}, and higher than the RSG \mdot\, rates in the de Jager prescription. These objects, originally studied in the LMC \citep{van2005empirical} and chosen for their inferred strong IR excess emission, were found to have high mass-loss rates and are sometimes OH maser emitters, both indicators of dense circumstellar material (CSM). If dense enough, this CSM could potentially obscure the star at visual wavelengths, though there has been no formal definition for how extinguished a star must be to qualify as ``dust enshrouded" (see Section 2). 

The DE-RSG mass-loss prescription \citep{van2005empirical} has been employed in some stellar evolution models \citep[e.g.][]{ekstrom2012grids}, despite these dust-enshrouded stars being rare.  Even taking the number of originally claimed DE-RSGs\footnote{We report below that most of these originally claimed DE-RSGs are actually not dust enshrouded, making true DE-RSGs even more rare.}, they must represent a minority of RSGs and therefore must represent a small fraction of the time during the RSG phase. The rarity of the DE-RSGs is well illustrated in the LMC:  Above a luminosity of \logl=5.0  \cite[where the sample of cool supergiants is likely to be complete, see Section 3.3 within][]{davies2018humphreys} only 9 out of 73 cool supergiants are included in \citeauthor{van2005empirical}'s DE-RSG sample, while even fewer, 4 out of 73, are OH maser sources. Although RSGs in clusters show that \mdot\ increases as the star approaches SN, \citeauthor{beasor2020new} did not find any RSG to have an \mdot\ value as high as those in the \citet{van2005empirical} sample.  This raises an important issue:  does the DE-RSG phase last long enough (and is \mdot\ high enough) for this phase to significantly influence the star's resulting SN type?

Here we take another look at the DE-RSGs from the \citet{van2005empirical} sample, using photometric data from an updated LMC cool supergiant catalogue \citep{davies2018humphreys}. Using a more recent catalogue with high spatial resolution and high sensitivity data shows that {\it the putative DE-RSGs do not stand out as a distinct group in colour-magnitude diagrams}. Re-appraising the mass-loss rates of the DE-RSGs, we find that most of them have rather low  \mdot\ commensurate with normal RSGs in clusters. Moreover, the \citet{van2005empirical} prescription overestimates the mass-loss of these stars, representing more of an upper threshold than a representative fit through our sample. As we discuss, this undermines the motivation for using the enhanced DE-RSG mass-loss rate prescription for any significant fraction of the RSG phase in stellar evolution models.

\section{Defining a dust enshrouded red supergiant}\label{sec:define}
If the DE-RSG phase is a bona-fide evolutionary stage that all RSGs will pass through, then it is useful to formally define this phase. It has been shown that quiescent mass-loss during the RSG phase is low \citep{beasor2016evolution,beasor2018evolution, beasor2020new} and is unable to remove more than $\sim$ 1\msun\ of material. As a result, for single-star evolution, the burden falls to hypothetical rare high mass-loss phases in order to peel away a large enough proportion of the hydrogen envelope to form a stripped star in the 20--30\msun\ mass range. Without this, an RSG cannot lose enough envelope to return to the blue of the Hertzsprung-Russel diagram (HRD). To achieve this through a high \mdot\ phase, a star would need to be experiencing mass-loss rates of $10^{-4}$\msun yr$^{-1}$\citep[higher than the quiescent \mdot, shown to be $\sim$ $10^{-6}$ -- $10^{-5}$\msun yr$^{-1}$, see e.g.][]{beasor2020new} for around 10\% of the 10$^{6}$ yr RSG phase \citep{georgy2013grids}. This would strip $\sim$ 10\msun of material, and allow the star to return to the blue of the HRD and die as a H-poor supernova \citep[e.g.][]{chieffi2013pre}. 

\begin{figure}
    \centering
    \includegraphics[width=\columnwidth]{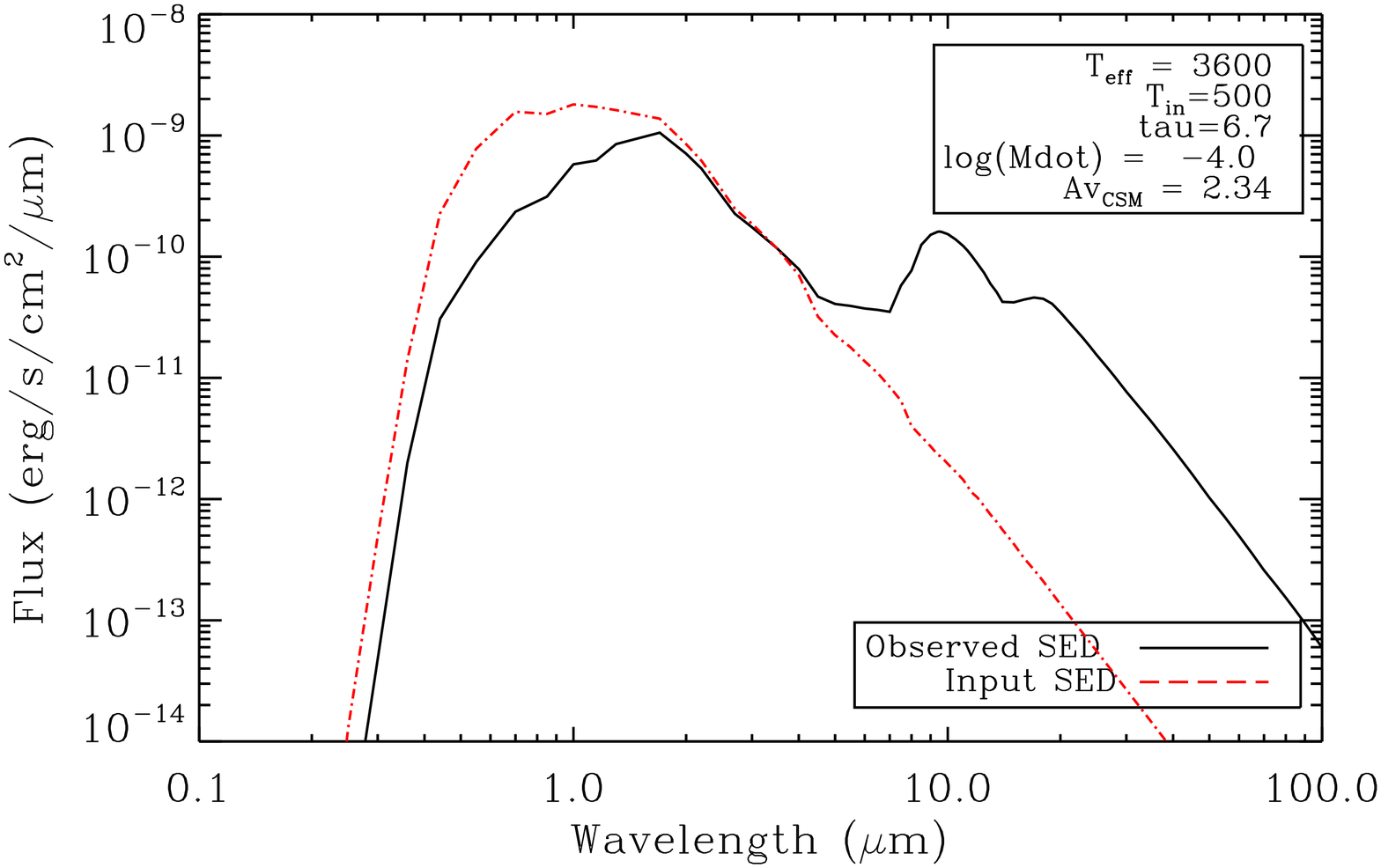}
    \caption{A model SED for a dust-enshrouded RSG. The solid black line shows the output spectrum, while the red dashed line shows the input SED.}
    \label{fig:dersg_def}
\end{figure}

In Fig. \ref{fig:dersg_def} we show a model SED from {\tt DUSTY} for a star where \mdot= 10$^{-4}$\msun\ yr$^{-1}$. The solid black line represents the observed SED, while the dashed red line shows the input SED (i.e. the intrinsic photospheric emission of the RSG). We can see that a high \mdot\ has a significant effect on the observed SED, with a strong silicate feature in emission at longer wavelengths ($\lambda$$>$10\micron) and a clear loss of flux at shorter wavelengths, corresponding to an optical extinction of $A_{\rm V}$ = 2.34 mag. Importantly, when reddening is due to mass-loss, any flux lost is re-emitted at longer wavelengths. Therefore, any DE-RSG candidate should have high optical extinction ($A_{\rm V} \geq 2$) and corresponding excess emission at $\lambda \geq$ 2$\mu$m, in addition to an \mdot\ exceeding $10^{-4}$\msun) yr$^{-1}$.

\section{Samples}\label{sec:sample}
The high \mdot\ sample of DE-RSGs from \citet{van2005empirical}, hereafter RSG-vL, were primarily assembled from samples of mid-IR sources with known optical counterparts \citep[e.g. bright IRAS sources, see][]{zijlstra1996obscured}, as well as bright mid-IR sources associated with heavily dust enshrouded objects \citep{zijlstra1996obscured}. We list these 14 objects in Table \ref{tab:results}. This sample of RSGs has formed the basis of a widely used \mdot-prescription, as we discuss in detail later. 

For this work, we compare the \citet{van2005empirical} sample to the LMC cool supergiant catalogue from \citet{davies2018humphreys}, hereafter RSG-DCB, which compiles RSGs from many studies \citep{elias1985m,levesque2006effective,gonzalez2015new,bonanos2009spitzer,buchanan2006spitz, van2005dusten,goldman2017wind}. Each cool supergiant (CSG) in the sample is confirmed either via SED shape or by spectroscopy, with 187 objects identified as spectral type M. The authors state that the sample is considered to be complete above luminosities of \logl = 5, and when looking at the HR diagram (Fig. 5 within) it appears that there is little incompleteness above \logl = 4.7. \citet{davies2018humphreys} calculate the luminosities by integrating under the spectral energy distribution (SED) of each star, under the assumption that any flux lost at shorter wavelengths is re-emitted at longer wavelengths.  We use those integrated luminosities in this work.

\begin{figure}
    \centering
    \includegraphics[width=\columnwidth]{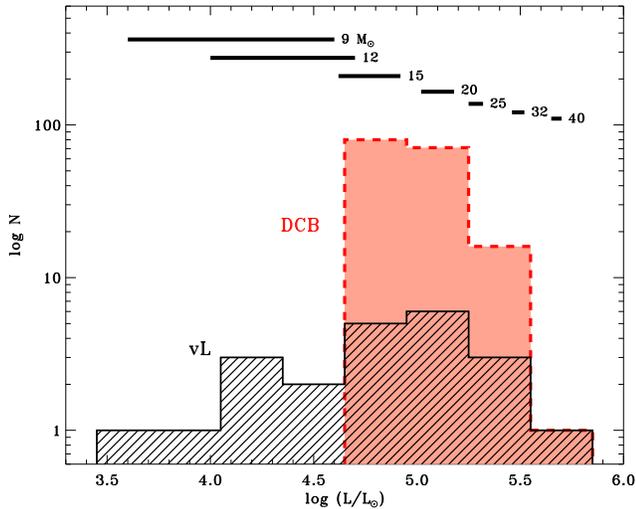}
    \caption{Histogram of observationally inferred luminosities for two samples of RSGs in the LMC.  The black hatched histogram (vL) is composed of the 22 dust enshrouded stars of M-type from van Loon et al.  
    The filled red histogram 
    (RSG-DCB) is a statistically complete catalogue of LMC CSGs with \logl\ $>$ 4.8. The black horizontal bars at the top show the range of luminosity during the RSG phase in stellar evolution models \citep{ekstrom2012grids} for a range of initial masses from 9 to 40 $M_{\odot}$. }
    \label{fig:ldist}
\end{figure}

In Fig.~\ref{fig:ldist} we show a histogram of the luminosities of the objects in the RSG-vL and RSG-DCB samples. Since the original sample from \citet{van2005empirical} also contained lower mass AGB stars, we only include RSG-DCB objects with luminosities above \logl = 4.6. But we caution that the precise cut-off between high mass AGB and low mass RSGs is difficult to define, so it is possible there is still remaining contamination (in fact, later we conclude that four of these are likely AGB stars).  

\section{Modelling the dust shells}\label{sec:dustshell}
We now take a closer look at the SEDs of the vL05 sample of DE-RSGs. For this, we employ the grid of {\tt DUSTY} models from \citet{beasor2016evolution,beasor2018evolution,beasor2020new}, see papers for full details\footnote{Note that modeling the SED of WOH G64 required a grid with higher optical depths.}. {\tt DUSTY} \citep[][]{ivezic1999dusty} is a code that solves the radiative transfer equation for a star surrounded by a spherically symmetric dust shell of a given optical depth ($\tau_{\rm V}$, optical depth at 0.55$\mu$m), inner dust temperature ($T_{\rm in}$) at the innermost radius of the dust shell ($R_{\rm in}$), and radial density profile ($\rho_{\rm r}$). We now briefly describe the dust shell models.

{\tt DUSTY} requires an input SED of a given effective temperature, \teff, for the initial star that is not obscured by dust. While the temperature scale for RSGs is debated \citep{levesque2005effective,davies2013temperatures,tabernero2018lte} it is generally agreed that RSGs are somewhere between 3400 - 4400K. We therefore repeat our fitting procedure (see Section \ref{sec:fitting}) using SEDs at temperatures of 3400K, 3600K, 3900K and 4200K.

As light leaves a star surrounded by a dust shell, it is absorbed by dust grains, and that absorbed luminosity is then re-emitted at longer wavelengths.  Different dust compositions change the output SED of the star. RSGs are known to have relatively large grains composed of silicate dust \citep[ this chemical composition is known to cause the characteristic 10 $\mu$m bump, e.g.][]{woolf1969silicate}. Therefore, we opt for the silicate dust described by \citet{draine1984optical}, and a grain size of 0.3 $\mu$m \citep[e.g.][]{scicluna2015large}. We also assume a gas-to-dust mass ratio of $r_{\rm gd}$=500, appropriate for LMC metallicity\footnote{The precise $r_{\rm gd}$ is uncertain, but recent studies have shown the range does not strongly deviate from 500 \citep[$r_{\rm gd}$$\sim$ 340 - 540][]{roman2014dust,goldman2017wind} } \citep[e.g.][]{van2005empirical}, and a grain bulk density of 3 g cm$^{-3}$. We note that $r_{\rm gd}$ is not well known and as such derived \mdot\ values could be scaled up or down accordingly, but this is beyond the scope of this work. As discussed in \citet{beasor2020new} the derived \mdot s will not be significantly higher than we find here unless one wishes to adopt a large non-standard $r_{\rm gd}$ value, for which there is currently no strong evidence.

To compute \mdot\ from the dust shell models we must also assume a wind speed ($v_{\infty}$) and a density distribution ($\rho_{\rm r}$). As we do not have wind speed measurements for each RSG in the sample, we instead use measurements from other RSGs \citep[e.g.][]{van2001circumstellar,goldman2017wind} and where there is no measured wind speed, we assume $v_{\infty}$ =  25$\pm$5 km s$^{-1}$. LI-LMC 4, LI-LMC 62, LI-LMC 77 and LI-LMC 92 all have measured wind speeds ($\sim$ 10 km s$^{-1}$) from maser emission in the literature \citep{goldman2017wind} and so for these objects we use their measured values. This is, however, a very low wind speed for a RSG \citep[which are generally higher, around 25 km s$^{-1}$, see e.g.][]{marshall2004asymptotic,van1999mass}, suggesting instead that these are likely to be luminous AGB stars and not RSGs.  This is discussed in more detail below along with other evidence that these four are indeed AGB stars. We assume a steady state wind with density falling off as $r^{-2}$. Note that the choice of steady state wind differs from the assumptions made in VL05, see discussion in Section \ref{sec:paramcomp}.

Combining all of this, the total \mdot\ can be calculated from 

 \begin{equation} \dot{M} = \frac{16\pi}{3} \frac{R_{in} \, \tau_{V}  \, \rho_d \, a \, v_\infty}{Q_{V}} \, r_{\rm gd}
 \end{equation}

\noindent where $Q_V$ is the extinction efficiency of the dust \citep[as defined by the dust grain composition,][]{draine1984optical} and $a$ is the grain size. The mass-loss rate is scaled with luminosity via the inner dust radius parameter, $R_{\rm in}$. The value for $R_{\rm in}$ is directly output by DUSTY, scaled to a standard \lbol\ of 10$^{4}$$L_\odot$. We then scale the value of $R_{\rm in}$ to the correct luminosity via 

\begin{equation} R_{\rm in} = (10^{L_{\rm bol}} / 1 \times 10^4)^{0.5},
\end{equation}

see DUSTY manual Section 4.1 for further explanation of this. We note that this differs from the scaling used for radiatively driven winds, where $R_{\rm in}$ $\propto$ $L^{0.75}$, see Section \ref{sec:paramcomp} for further discussion on the appropriateness of assuming radiatively driven winds for RSGs.

\subsection{Departures from spherical symmetry}
While it has been shown that the dust shells of evolved stars are clumpy \citep[e.g.][]{decin2020sub}, here we model all the stars with spherically symmetric dust shells. We note that while asymmetric envelopes could lead a model to underestimate bolometric luminosities, since some starlight escapes along lines of sight with low dust optical depth, asymmetry will not significantly compromise the derived mass of emitting dust or the derived \mdot\ unless \lbol\ is vastly over or underestimated. They will however, influence the \mdot\ prescription expressed as a function of luminosity. 

In cases where \lbol\ is significantly over or under-estimated, the derived \mdot\ would be affected. As explained above, the final value of \mdot\ depends on the luminosity of the star through the parameter $R_{\rm in}$. For stars with highly clumped CSM, the line-of-sight column density to the star may be lower than expected from spherical symmetry. In other words, there is a large amount of CSM but it is not significantly obscuring the star. This can be identified from a predicted optical SED that is far fainter than the observed optical SED. In this case, by simply integrating under the SED the calculated \lbol\ would be over estimated, leading to an over estimate of \mdot. WOH G64 is a clear example of this. Due to it's dusty torus, when deriving \lbol\ by integrating under the observed SED the luminosity is overestimated \citep[\logl=5.8,][]{davies2018humphreys}. When taking into account the true geometry of WOH G64's CSM, the derived \lbol\ is reduced \citep[\logl=5.45,][]{ohnaka2008spatially}.

In the reverse case, if the model SED overestimates the optical flux this could be due to highly clumped CSM sat directly in front of the star. In this case, integrating under the SED gives an underestimate of \lbol\ and hence of \mdot. For the stars analysed in this work, we do not see any cases of this occurring for stars with \lbol$\geq$5. This effect may however be able to explain the poorer model fits to LI-LMC 60, LI-LMC 77 and LI-LMC 92, but as these are all low luminosity (\lbol$\leq$ 4.60) and likely AGB stars (see Section \ref{sec:results}) the conclusions of the paper are unaffected.

\subsection{Comparison to van Loon et al.\ models}\label{sec:paramcomp}
We now compare our parameter choices to those made by \citeauthor{van2005empirical}, and discuss what impacts these differences may have on the calculated \mdot. 

Both works assume a constant fiducial grain size (0.1 $\mu$m in VL05 and 0.3 $\mu$m here). However, in VL05 it is stated that this grain size ($a$) is altered if the fit of the {\tt DUSTY} model to the observed SED is significantly improved by a different value. Consequently, the authors also use smaller constant grain sizes (0.05 - 0.06 $\mu$m) for 3 objects included in this work (LI-LMC 4, LI-LMC 60 and WOH G64), as well as the MRN grain distribution (a power law with an exponent of -3.5) with minimum and maximum grain sizes at 0.01 and 0.1 $\mu$m, respectively. More recent observations of the dust surrounding RSGs shows that the average grain size is likely far larger, with the dust surrounding VY Canis Majoris having an average diameter of 0.5 $\mu$m \citep[e.g.][]{scicluna2015large,smith2001asymmetric}. The MRN distrubution is more commonly associated with interstellar dust grains \citep{mathis1977size} and has not been shown to be appropriate for RSGs. In \citet{beasor2016evolution} the impact of varying grain size on the measured \mdot\ value was investigated, by creating 6 different grids of {\tt DUSTY} models, examining the effect of the MRN distribution and constant grain sizes between 0.1 - 0.5 $\mu$m in steps of 0.1 $\mu$m. While grain size had a marginal effect on extinction, the total \mdot\ was unaffected by varying grain size (see Fig. 9 within). 

We also differ in our choice for assumed density profiles within {\tt DUSTY}. Here, we choose to use a steady-state wind where density falls off as $r^{-2}$, whereas \citet{van2005empirical} use the analytic approximation for radiatively driven winds. Under the radiatively driven wind scenario, the acceleration zone for the dust is slower, and hence a higher density of dust lies closer to the star.   The net effect of this is that for a given \mdot, the dust lies closer to the star and less mass is required to make a stronger mid-IR excess. However, the driving mechanism for RSG winds is uncertain. It was initially thought that RSG winds were driven by the same mechanism as AGB stars, radiative pressure on dust grains \citep[e.g.][]{gehrz1971mass}, but the higher effective temperatures and lower pulsational amplitudes of RSGs make this scenario unlikely. Combined, these effects would require the dust to reach a larger height proportionally to the stellar radius, and less dust may be levitated to begin with \citep[see e.g.][]{ohnaka2008spatially,arroyo2015extension, kee2021analytic}. We therefore model the RSG wind as a steady state wind, making assumptions about the density distribution and outflow velocity. Note that this different choice would tend to make \mdot\ values higher in our study, whereas we actually find slightly lower \mdot\ values as compared to van Loon et al.

\section{Model fitting}\label{sec:fitting}
To find a best fit model for each star, we compute a grid of {\tt DUSTY} models varying in $T_{\rm in}$ and $\tau_{\rm V}$ for each input $T_{\rm eff}$. In each grid, $\tau_{\rm V}$ is varied between 0 - 10 in steps of 0.1 and $T_{\rm in}$ spanning 100 - 1200K in steps of 100K. For each model, we compute synthetic photometry to match the observed photometry, taking filter profiles from SVO Filter Profile Service\footnote{http://svo2.cab.inta-csic.es/theory/fps/} and convolving with the SED. 

We use $\chi^{2}$-minimisation 

\begin{equation}
\chi^2 =\sum_i \frac{ (O_{i}-E_{i})^2 }{\sigma_i^2}
\end{equation}
where $O$ is the observed photometry, $E$ is the model photometry, $\sigma^{2}$ is the error and $i$ denotes the filter. The best fit model is that which produces the lowest value of $\chi^{2}$. Since the errors in this study are primarily driven by systematics (e.g. SED temperature, extinction law) it is not appropriate to assume Gaussian errors. Instead, we define the errors as the minimum $\chi^2$+10 \citep[see][for further discussion]{beasor2020new}.

\subsection{Results}\label{sec:results}
For each RSG in the vL05 sample, we derived \mdot\ values using the photometry presented in the \citet{davies2018humphreys} catalogue\footnote{The photometry in the \citet{davies2018humphreys} catalogue, including WISE and Spitzer data, differs from that used by \citet{van2005empirical} (ISO, IRAS, Spitzer-MIPS). This may account for some of the difference in results.}. Since foreground extinction is present for each RSG, we de-redden the photometry using the LMC extinction law from \citet{gordon2003quantative} and the corresponding $A_{\rm V}$ value for each RSG \citep[from the extinction maps of][]{zaritsky2004magellanic}, also presented in \citet{davies2018humphreys}.

\begin{figure}
     \centering
     \includegraphics[width=\columnwidth]{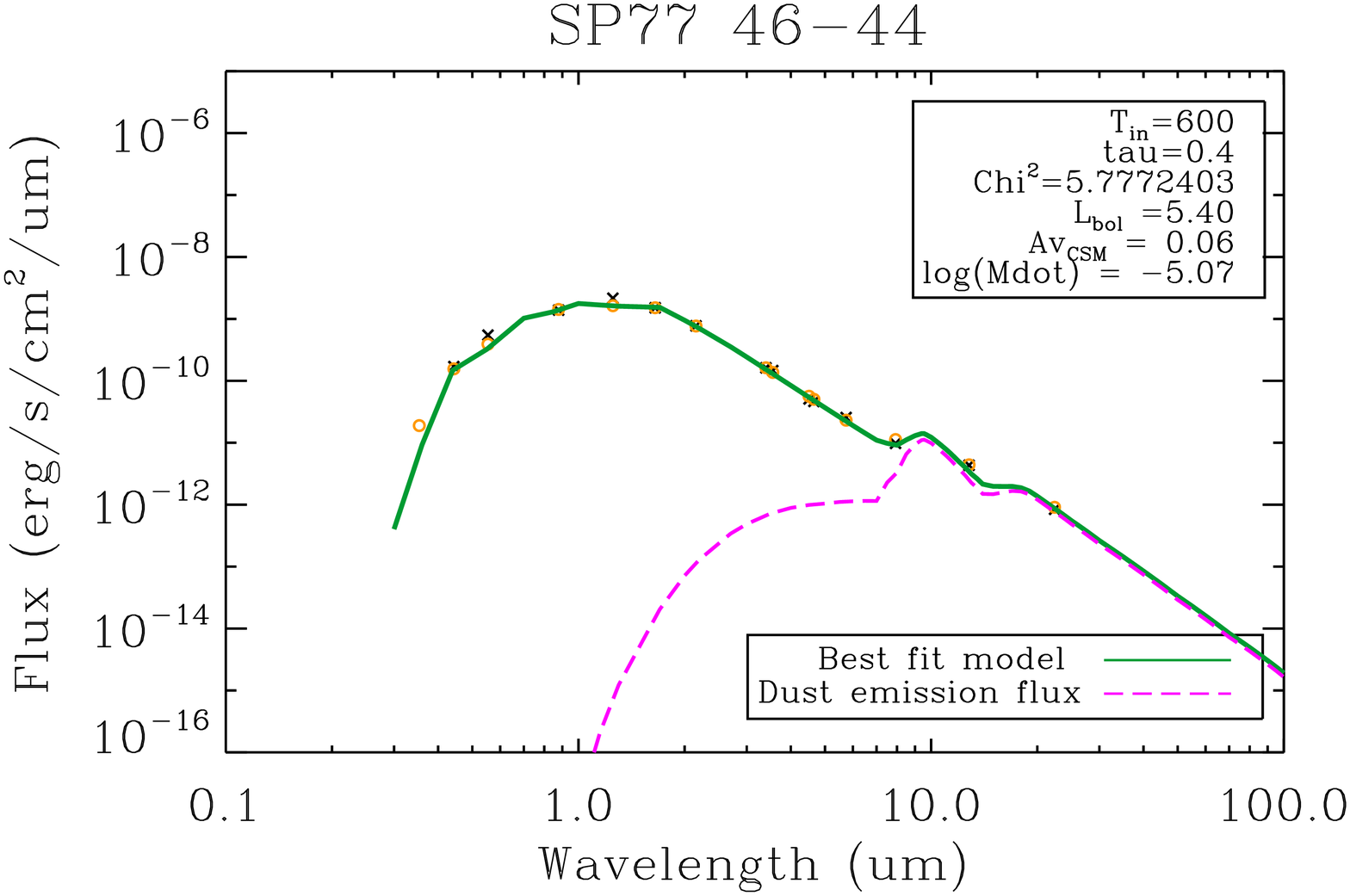}
     \caption{The best fit model for SP77 46-44. The solid green line shows the best model fit from this work and the pink dashed line shows the flux from dust emission. }
     \label{fig:dustshell1}
 \end{figure}

The \mdot\ values that we calculated for each star in the vL05 sample are provided in Table \ref{tab:results}. Note that in this work we do not determine luminosity from {\tt DUSTY} models, instead we use the values presented in \citet{davies2018humphreys}, which are estimated by integrating under the observed SED under the assumption that any light lost at shorter wavelengths is re-emitted at longer wavelengths. While the photometry presented in the \citet{davies2018humphreys} catalogue is not contemporaneous, it is unlikely this will have a significant effect on the derived \lbol\ and hence \mdot. In extreme cases, RSGs vary in the $V$, $R$, $K$ and mid-IR with amplitudes of around 1, 0.5, 0.25 and $<$0.1 mag respectively \citep{levesque2007late,yang2018red,ren2019period}. It is not clear if RSG variability is bolometric (i.e. a change in the intrinsic luminosity) or due to a change in temperature, but if we assume that variability {\it is} caused by a change in overall luminosity at most we would expect this to be around 0.2 dex. This level of variability would not have a significant effect on the derived \mdot\ \citep[see also][]{beasor2018thorne,beasor2021wd1}. We note, however, that the luminosities determined using {\tt DUSTY} are on average only 0.04 dex different from the \citeauthor{davies2018humphreys} results, with the exception of WOH G64 which is 0.3 dex fainter. This is likely due to modelling the dust as a spherically symmetric shell, whereas WOH G64 is known to be highly assymetric \citep{ohnaka2008spatially}.  Figure \ref{fig:dustshell1} shows an example best-fit model for SP77 46-44, which is a more typical example. The plot shows contributions to the output spectrum, including the dust emission flux (pink dashed line). In the appendix we show the best fits for all other RSGs in the vL05 sample.

In Table \ref{tab:results} we also show the previously found best-fit \mdot\ values from vL05. In all cases, we find the best fit \mdot\ is slightly lower using the photometry from the updated \citet{davies2018humphreys} catalogue, but still consistent to within the errors. 

\begin{table*}[]
    \centering
    \begin{tabular}{l l c c c c c c }
    \hline
      \multicolumn{2}{c}{ID}   & $log(L/L_{\rm \odot})$ & \teff & $T_{\rm in}  $ & $\tau_{\rm V}$ & \multicolumn{2}{l}{log(\mdot/M$_{\rm \odot}$ $yr^{-1}$)} \\
      DCB18 & vL05 & DCB18 & K & K & & This work  & vL05  \\
      \hline
     WOH G 64 &$-$&5.45$^\dagger$ &3400&$600^{+400}_{-100}$&17$^{+3}_{-1}$&$-3.13^{+ 0.28}_{- 0.41}$& -3.12 \\
  HV 12501 &$-$&5.19&3400&$ 500^{+ 200}_{- 100}$&$0.50^{+0.10}_{-0.20}$&$-4.94^{+ 0.17}_{- 0.18}$& -4.79\\
  HV 2360 &-&5.15&3400&$ 600^{+ 200}_{- 200}$&$1.00^{+0.30}_{-0.20}$&$-4.76^{+ 0.24}_{- 0.16}$& -4.51\\
  HV 2561&-&5.29&3400&$ 500^{+ 200}_{- 100}$&$0.60^{+0.10}_{-0.40}$&$-4.81^{+ 0.17}_{- 0.36}$& -4.47 \\
  HV 5870&-&5.02&3400&$ 400^{+ 100}_{-   0}$&$0.90^{+0.20}_{-0.10}$&$-4.53^{+ 0.05}_{- 0.17}$& -4.68 \\
  HV 888 &-&5.48&3400&$ 600^{+ 300}_{- 100}$&$0.80^{+9.20}_{-0.20}$&$-4.70^{+ 1.03}_{- 0.20}$& -4.43 \\
  HV 916&-&5.27&3400&$ 600^{+ 200}_{- 100}$&$1.00^{+0.30}_{-0.10}$&$-4.70^{+ 0.17}_{- 0.18}$& -4.50\\
  HV 986&-&5.10&3400&$ 300^{+ 100}_{- 200}$&$0.30^{+0.00}_{-0.20}$&$-4.99^{+ 0.23}_{- 0.25}$& -4.89\\
  HV 996&-&5.13&3400&$1200^{+   0}_{-   0}$&$3.00^{+0.30}_{-0.20}$&$-4.75^{+ 0.05}_{- 0.04}$& -4.28\\
LI-LMC 4&IRAS 04407-7000 &4.63&2500&$1100^{+ 100}_{- 100}$&$7.00^{+3.00}_{-1.00}$&$-4.50^{+ 0.09}_{- 0.10}$& -4.15\\
LI-LMC 60&IRAS 04498-6842 &5.04&2500&$1000^{+ 200}_{- 100}$&$3.40^{+0.50}_{-0.40}$&$-4.60^{+ 0.10}_{- 0.13}$& -4.30\\
LI-LMC 77&IRAS 04509-6922 &4.80&2500&$ 900^{+ 100}_{- 100}$&$3.80^{+0.40}_{-0.40}$&$-4.57^{+ 0.10}_{- 0.11}$& -4.45\\
LI-LMC 92&IRAS 04516-6902 &4.90&2500&$ 600^{+ 100}_{- 100}$&$4.20^{+0.50}_{-0.50}$&$-4.17^{+ 0.13}_{- 0.18}$& -4.23 \\
SP77 46-44&-&5.40&3400&$ 600^{+ 400}_{- 200}$&$0.40^{+0.10}_{-0.20}$&$-5.07^{+ 0.16}_{- 0.44}$& -4.56\\ \hline
\multicolumn{8}{l}{$^\dagger$Due to the non-spherical morphology of the dust shell, the luminosity for WOH G64 is taken from \citet{ohnaka2008spatially}.}
    \end{tabular}
    \caption{Results from DUSTY fitting for the RSGs in the vL05 sample. We also show the \mdot\ values presented in vL05.}
    \label{tab:results}
\end{table*}

\begin{figure*}
    \centering
    \includegraphics[width=12cm]{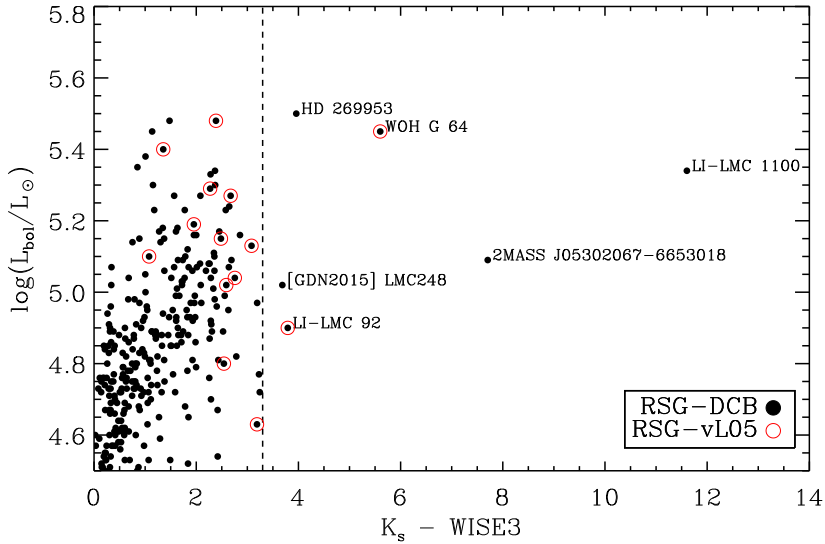}
    \caption{Colour luminosity diagram used to identify candidate DE-RSGs. The red outlines show the objects originally included in the vL05 sample, while the black dots show the objects in the DCB18 LMC catalogue.}
    \label{fig:cmd_vL}
\end{figure*}

For most of the objects in this sample, we are able to achieve good fits across the full SED. However for LI-LMC 4, LI-LMC 60, LI-LMC 77 and LI-LMC 92 we are unable to accurately match the fluxes at wavelengths shorter than 2$\mu$m with effective temperatures above 3400K. We instead found the best fitting results (i.e. lowest $\chi^2$ values) when using a 2500K SED, which is lower than the expected temperature range for RSGs \citep{levesque2005effective,davies2013temperatures,tabernero2018lte}. Given their low luminosities (\logl $\leq$ 5), large pulsational amplitudes (d$K_{\rm s}>$1.2 mag) and long pulsation periods ($>$ 1000 days) it is likely that they are instead luminous AGB stars. Indeed, they are listed as being Mira type variables by \citet{groenewegen2018luminosities}. We note however that despite difficulty matching the shorter wavelengths, the objects do not appear to have significant mid-IR excess (see Fig. \ref{fig:cmd_vL} and \ref{fig:appp2}).% and are undergoing moderate mass-loss (\mdot$<10^{-4}$\msun) yr$^{-1}$). We also note that these objects have lower wind speeds than typical RSGs (see Section 4), suggesting a lower escape velocity and a lower mass star for the same luminosity and stellar radius.

We do note, however, that the Spitzer IRAC1 \& 2 (3.5 and 4.5$\mu$m) photometry appears discrepant with the WISE 1 \& 2 photometry (3.4 and 4.6$\mu$m) for these 4 stars. While it is unclear the cause of this discrepancy, by repeating the model fitting process while omitting either the Spitzer or WISE data we found the effect on the derived \mdot\ was less than 5\%. While the stars do appear reddened, we stress that is not possible for this to be due to mass-loss, since it is only possible in extreme circumstances to lose significant flux at shorter wavelengths due to dust absorption without corresponding excess emission at longer wavelengths \citep[see Fig. \ref{fig:dersg_def} and ][]{kochanek2012absorption}.  

%slower Vs, AGB wind model, scale 

\section{Discussion}
Fitting the SEDs using {\tt DUSTY} reveals that the vast majority of RSGs in the vL05 sample (all except one) have weak mid-IR excess emission and low \mdot\ values commensurate with normal RSGs, and low levels of circumstellar extinction that do not make them seem particularly enshrouded in dust. 
The one exception to this is WOH G64, which is known to have an unusual morphology, and for which we determine a high mass-loss rate of log(\mdot/$M_{\odot}$) = $-$3.13. The circumstellar extinction from WOH G64's dust shell  ($A_{\rm V}$ = 7.92 mag) is also significantly higher than any other objects in our sample.  

Below, we also investigate the DCB sample further to determine if there are any additional DE-RSGs based on color selection that might have been missed in the original RSG-vL sample.

%\begin{figure}
%    \centering
%    \includegraphics[width=\columnwidth]{mdotcomp_oct.eps}
%    \caption{Comparison of the mass-loss rates derived here and in VL05, in units of log(\mdot/\msunyr{}). The dashed line shows the 1:1 correlation. We find the \mdot\ values presented here are on average 1.3 dex lower than those presented in VL05.  }
%    \label{fig:comparevl}
%\end{figure}

\subsection{Identifying new dust enshrouded candidates}\label{sec:separations}
We now use observational tracers and the \citeauthor{davies2018humphreys} catalogue to search for other candidate DE-RSGs in the LMC. To do this, we compare the colour magnitude diagrams for the CSGs in the DCB sample and the vL05 sample. As dust accumulates around a star, this dust absorbs flux at shorter, optical wavelengths, and then re-emits this flux at longer wavelengths \citep[e.g.][]{gehrz1971mass}. This causes the stars to appear reddened at visual wavelengths and to have mid-IR excess, yielding a very red optical/IR color. We use [K-WISE3] as a tracer for dust (i.e. redness) and \lbol\ from \citep{davies2018humphreys} for each of the stars in the sample. We point out the location of the well-known dust enshrouded star WOH G64, a star which has an extreme morphology \citep[e.g.][]{ohnaka2008spatially}, and here has again been shown to have both high circumstellar extinction and a high mass-loss rate.

Figure \ref{fig:cmd_vL} shows the position of the stars from the DCB sample (black circles) and the vL05 sample (red circles) in colour-luminosity space. The $K_{\rm S}$-WISE3 colour is chosen as it provides a long lever arm with which to characterise the mid-IR excess, probing both photospheric emission ($K_{\rm S}$) and dust emission (WISE3). We can see that there is no clear distinction between the normal RSG population of the LMC or the majority of the putatively dust-enshrouded RSGs from the vL05 sample. We identify only 5 additional stars (besides WOH G64) with high luminosities (i.e. likely RSGs and not AGB stars) and $K_{\rm S}$-WISE3 colours redder than 3.4. We will now discuss each of these objects in more detail.

\subsubsection{LI-LMC 1100}
LI-LMC 1100 (also known as  IRAS 05280-6910) is an extremely bright mid-IR and OH/IR maser object in the LMC cluster NGC 1984 \citep{wood1992asy,van2001circumstellar}. As of yet, the object does not have a derived spectral type. \citet{asad2017young} estimate the age of the cluster to be 7 -- 10 Myr from the integrated light, suggesting that LI-LMC 1100's initial mass was roughly 20 -- 25\msun. An age of 7 - 10 Myr is too old for a very luminous embedded protostar, but it is just right for a very luminous RSG.  ATCA observations located the source to be very near to another RSG WOH G 347, but L' band imaging revealed two separate sources $\leq$5'' apart, with WOH G347 contributing only 5 per cent to the total emission in IRAS \citep{van2005dusten}. Consequently, LI-LMC 1100 appears to be the most reddened object in the LMC, far exceeding the redness of the famous dust enshrouded RSG WOH G64 (see Fig.~\ref{fig:cmd_vL}), which is thought to be enshrouded by a thick dusty torus \citep[e.g.][]{ohnaka2008spatially}. 

\citet{goldman2017wind} discuss LI-LMC 1100 in some detail, stating that the objects' unusual polarisation and complex maser peaks imply it is surrounded by a complex envelope, indicative of extreme mass-loss. Indeed, VY CMa shows a similary distinct maser profile (e.g. Cohen et al. 1987) and the morphology of VY CMa's CSM is known to be complex, consisting of many arcs and knots \citep[e.g.][]{monnier1999gasps,smith2001asymmetric,smith2009red,smith2004potassium,decin16}.  Due to its unusual morphology and a lack of clear photometric data in the mid-IR, it was not possible for Goldman et al. to derive a reliable \mdot\ value for LI-LMC 1100. 

\begin{figure}
    \centering
    \includegraphics[width=\columnwidth]{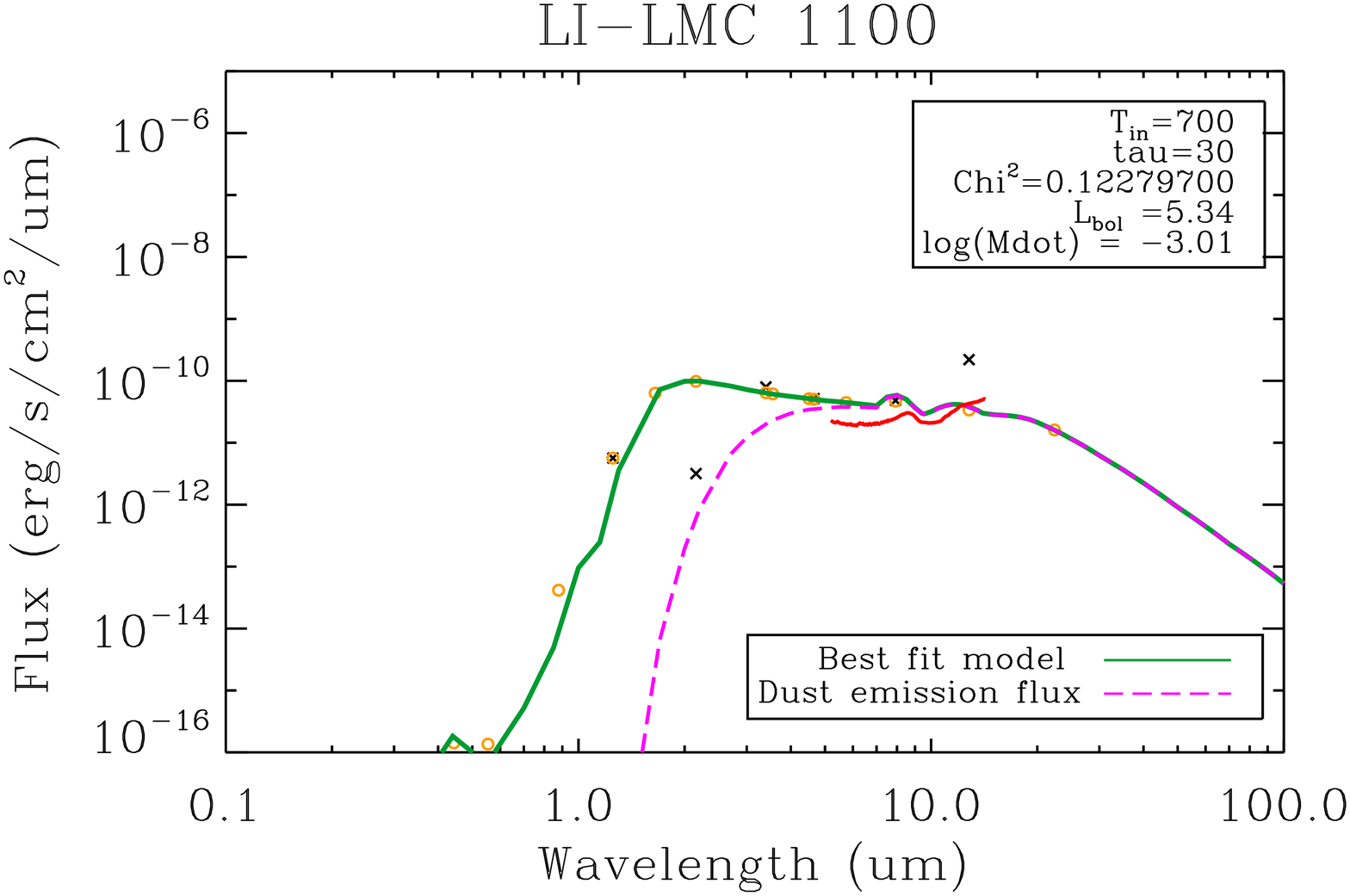}
    \caption{Best fit model for LI-LMC 1100, where the black crosses represent observed photometry, yellow circles represent synthetic photometry, the solid green line shows the best fit {\tt DUSTY} model and the pink dashed line shows corresponding dust emission. We also show a Spitzer-IRS spectrum (solid red line) though we do not use this for fitting. sWe note that despite the low $\chi^2$ value, two photometric points are clearly discrepant with the best fit model. It is clear however that if LI-LMC 1100 is an RSG, it is highly extinguished, with no visible flux below 1$\mu$m.}
    \label{fig:LILMC1100}
\end{figure}

Nevertheless, we fit the observed SED of LI-LMC 1100 using a coarse {\tt DUSTY} grid where $\tau_{\rm V}$ is between 10 and 30 in steps of 1, and show our best fit SED for the object in Fig. \ref{fig:LILMC1100}. We note that while the spherical models almost certainly do not accurately characterise the objects' luminosity, the total dust shell mass should still be able to be retrieved even if is is strongly asymmetric \citep[see Section 2.1.5 within]{beasor2016evolution}. Confirmation of the nature of LI-LMC 1100 requires further data to be taken, such as high resolution imaging to measure the flux without contamination from WOH G347, or infrared spectroscopy to confirm the object as an RSG. 

\subsubsection{HD 269953}
\citet{davies2018humphreys} list the spectral type of HD 269953 as G0 and find \logl=5.50, making this object a yellow supergiant. While it is not an RSG, the colours indicate it is likely a highly reddened. Follow up study would reveal whether this is an extreme object similar to IRC+10420. 

\subsubsection{[GDN2015] LMC 248}
While this object appears red in colour-magnitude space, the SED fitting returns a best fit \mdot\ of 10$^{-5}$ M$_{\odot}$ yr$^{-1}$. Inspection of the SED shows very little optical extinction caused by CSM, and therefore we do not consider LMC 248 to be dust enshrouded. It may appear slightly reddened due to temperature (it is an M3.5 spectral type) or line-of-sight ISM foreground extinction and reddening. 

\subsubsection{LI-LMC 92}
Our best fit model for this object finds \mdot\ = 10$^{-5.48}$ M$_{\odot}$ yr$^{-1}$, and inspection of the SED shows minimal flux loss at shorter wavelengths. We note that the lowest $T_{\rm eff}$ in our model grid is 3400K, and it was not possible to fit the entire SED reliably at this temperature. The SED of LI-LMC 92 likely requires a cooler input $T_{\rm eff}$, and since the temperature range for RSGs, while debated, is considered to be between 3400-4200K \citep{levesque2006effective,davies2013temperatures} we suggest this object is likely a luminous AGB star (recall that this is one of the four likely AGB stars discussed above in Section 5.2). 

\subsubsection{2MASS J05302067-6653018}
Upon inspection of high-resolution HST visual-wavelength images of this object, it is clear that 2MASS J05302067-6653018 is actually a cluster of stars and not an RSG. It is possible the object is comprised of several unresolved lower-mass red giants. 

\smallskip

Having considered all of the known luminous and red candidates for DE-RSGs in the LMC, we conclude that the only confirmed DE-RSG in the LMC is WOH G64, with LI-LMC 1100 being a strong candidate DE-RSG that would benefit from additional study.  The other luminous red objects have plausible explanations that are not DE-RSGs.

\subsection{A high mass-loss phase prior to SN}

In previous works using RSGs in clusters, we have shown that when a star arrives at the RSG phase it is undergoing only weak mass-loss  ($<$10$^{-6}$\msunyr{}), before evolving up the RSG branch toward SN. As the star does so, it increases in both luminosity and \mdot\ \citep{beasor2016evolution,beasor2018evolution,beasor2020new}. While \mdot\ does increase as the star moves up the RSG branch, quiescent RSG winds clearly do not climb steadily up to the levels previously estimated for DE-RSGs. \citeauthor{beasor2020new} showed that quiescent mass-loss during the RSG phase remains relatively low, with no RSG observed reaching the same extreme \mdot\ values as seen in the \citet{van2005empirical} sample {\it even for the most evolved RSGs.}

While some DE-RSGs do exist in the Milky Way and LMC, they are rare and must represent a short-lived or episodic phase of mass loss. But just how short is this phase, and when exactly does this episodic mass loss occur?  Is it limited to the very end of a star's life just before the SN?  As noted earlier, there is ample evidence from SNe with dense CSM for short-lived phases of pre-SN mass loss \citep{smith2014mass}.  Or instead, do those episodes of higher RSG mass loss occur intermittently during the RSG phase?  Or do they only occur for a subset of stars with special evolutionary paths (i.e. post-interaction binaries like merger products and common envelope phases, for example), or only for certain initial mass ranges?

The possibility of an enhanced \mdot\ phase immediately prior to SN was discussed by \citet{beasor2020new}, but since this phase would likely be short lived, it could be difficult to observe among the limited sample of RSGs in clusters. While quiescent mass-loss can likely only strip away $\sim$1\msun\ of material, a short-lived enhanced \mdot\ phase (where \mdot\ $>$10$^{-4}$\msun\ yr$^{-1}$) could have more impact on evolution if it has a long enough duration of significantly more than 10$^4$ yr.  

In this work, our re-appraisal of the \mdot s of the DE-RSGs from vL05 has shown that episodes of extreme mass-loss that may cause dust-enshrouded phases are far less common than previously thought.  We found only two candidates for DE-RSGs --- WOH G64 and LI-LMC 1100 --- amongst the full sample of 187 known RSGs in the LMC. From the original sample of 14 putatively dust enshrourded RSGs, all but one turned out to be normal RSGs with relatively low mass-loss rates. While the true nature of LI-LMC 1100 cannot be confirmed without further observation, we include both of these objects to estimate the fraction of DE-RSGs and to constrain their expected lifetime.  We find that the DE-RSGs represent only 3\% of the LMC RSG population. If these objects represent a bona-fide high \mdot\ phase that all single RSGs pass through, it would translate to only a tiny fraction of the RSG phase lasting only about 10,000 yr or less. Even extreme \mdot\ (on the order of that experienced by WOH G64) over this length of time would probably manage to rid the star of only $\sim$1--2\msun\ material, and would have little impact on the subsequent evolution, regardless of exactly when it occurs in the evolution. Of course, this line of reasoning overlooks the lack of a physically motivated explanation for a sudden and temporary superwind phase in single RSGs. On the other hand, if the very rare DE-RSGs correspond to special evolutionary histories like post-merger objects, then this DE-RSG phase might not be relevant for single RSGs at all.

\subsection{Optical reddening of RSGs from CSM}
\begin{figure*}
    \centering
    \includegraphics[width=18cm]{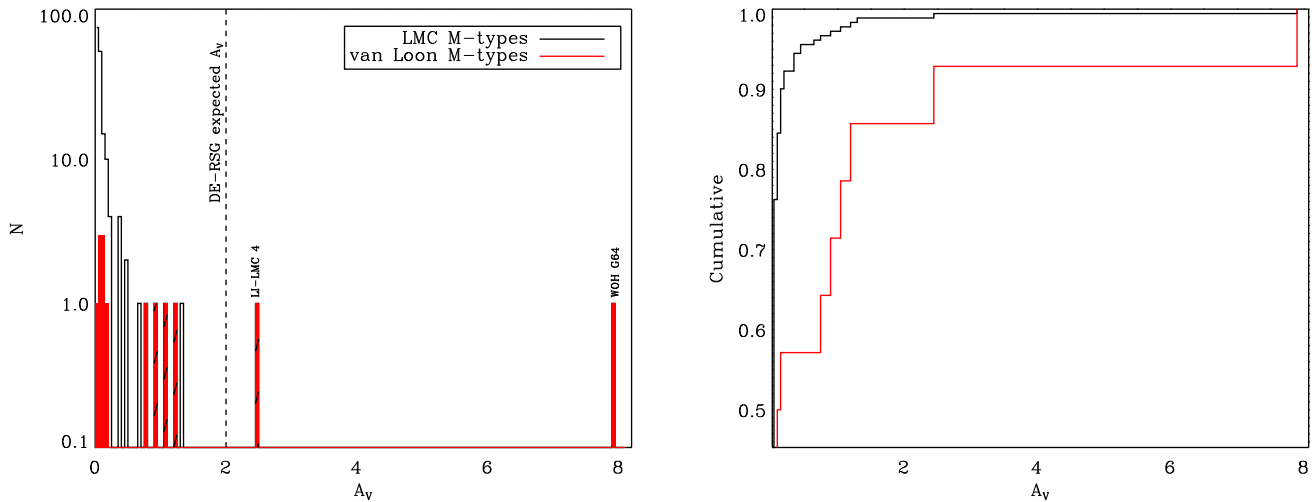}
    \caption{{\it Left:} Histogram showing the distributions of CSM $A_{\rm V}$ for the LMC M-type stars (black lines) and the vL05 M-type stars (red fill). The 4 lines shaded with horizontal black lines indicate the locations of the 4 AGB stars, LI-LMC 4, LI-LMC 60, LI-LMC 77 and LI-LMC 92. {\it Right:} Cumulative distribution for each sample. }
    \label{fig:av_plot}
\end{figure*}

The mass-loss undergone by the DE-RSGs leads to the build up of dust surrounding the star, which can potentially obscure the star at optical wavelengths. As defined in Section \ref{sec:define}, we expect a true DE-RSG to have optical CSM extinction of $A_{\rm V}$ $>$ 2.3 mag, corresponding to a factor of $\sim$10 reduction in its visual-wavelength flux. Previous works \citep[e.g.][]{massey2005smoke} have suggested that even at low \mdot\ values ($<$ 10$^{-8}$\msunyr) the optical extinction from CSM can be as high as $A_{\rm V}$ = 6 mag. This is not possible with steady, spherical winds with such low mass-loss rates, and would require a past episode of high mass loss. In Fig. \ref{fig:av_plot} we show the distribution of CSM $A_{\rm V}$ values for the DE-RSGs (red shaded region) and the LMC population of M-type stars (black solid line). From this plot, it is clear that the vast majority of RSGs {\it do not} appear significantly obscured by the CSM. Furthermore, the previously proposed DE-RSGs do not stand out as a unique group, and instead mostly have $A_{\rm V}$ values commensurate with the M-type population as a whole. This is also demonstrated by the right-hand plot of Fig. \ref{fig:av_plot}, which shows the cumulative distribution of extinction values for each sample. For example, around 40\% of the original vL05 sample and more than 90\% of the full RSG sample have no significant extinction at all (i.e. less than 0.2 mag of optical extinction). Comparing these samples yields a KS-test probability of 0.93, strongly implying that the vL05 sample and the rest of LMC RSGs are drawn from the same distribution. This is consistent with their distribution of various IR colours, which are also intermixed with normal RSGs. (We highlight the location of the 4 objects that are likely luminous AGB stars with diagonal black lines.) The one glaring exception is the extreme red object WOH G64, which we find to have an optical extinction of 7.9 mag, consistent with this star being the only {\it bona fide} DE-RSG in the LMC (and one candidate DE-RSG, LI-LMC 1100).

\subsection{Possible effects on the appearance of SNe}
While these high mass-loss phases are unlikely to cause single stars to end their lives as H-poor SNe, it could possibly change the appearance of the resulting H-rich SN. RSGs have been confirmed as the direct progenitors to Type II-P SN \citep[e.g.][]{smartt2009death,maund2005hubble}, but if extreme and short-lived \mdot\ episodes are a common precursor to SNe \citep{forster2018delay,morozova17} a major outburst would allow a significant mass of circumstellar material (CSM) to remain close to the star upon explosion. There are two important observational ramifications of this for SNe. First, SN progenitors surrounded by a compact shell might suffer additional extinction beyond the amount of interstellar line-of-sight extinction inferred from the SN itself (i.e.  using interstellar Na~{\sc i} D absorption in the SN spectrum), since this innermost CSM dust would likely be destroyed early.  If directly detected in pre-explosion archival data at visual wavelengths, such SN progenitors might have their luminosity and initial mass significantly underestimated.  Second,  the SN shock wave would then have to travel through this dense shell, changing the appearance of the SN, potentially allowing narrow H lines to form in the spectrum and therefore making it appear as a Type~IIn event for some time.  (Note that even if this final mass-loss phase can shed the entire H envelope in the last $\sim$1000 yr of the star's life, it is still unlikely to produce a SN~Ibc, because the H envelope that was just shed by a slow wind is still close enough to the star to make it appear as a Type~IIn due to the ensuing CSM interaction.)  It has been suggested that extreme dusty RSGs such as VY Canis Majoris (VY CMa) could be direct progenitors to Type~IIn SNe, rather than exploding as Type II-P events \citep{smith2009red,smith2009sn05ip}.

\subsection{Implementation in stellar evolutionary models}

Despite the vL05 prescription being derived solely from stars thought to be dust enshrouded and hence representing a specific phase in evolution, as opposed to being representative of the entire RSG population, the prescription has been implemented in a number of stellar evolutionary models \citep[e.g.][]{vanbev2007wolf,ekstrom2012grids, chieffi2013pre}. 

In each of these models, it is found that relatively low mass ($\sim$20\msun) single stars are able to shed their envelopes during the RSG phase, and evolve to the blue of the HR diagram where they will end their lives as Type Ibc SN. However, this blueward evolution results from the use of persistent enhanced mass-loss rates, which do not reflect the mass-loss rates of real RSGs.

It has been shown that quiescent mass-loss during the RSG phase is actually rather weak \citep{beasor2020new}, only stripping up to $\sim$1\msun\ of material prior to SN.  This is far below the threshold required to cause blue-ward evolution, because blueward evolution requires almost the entire H envelope to be shed. One might then speculate that an  enhanced,  high \mdot\ phase may exist, represented by the DE-RSGs, in which a prolonged outburst could cause a star to shed a high fraction of its envelope in a relatively short amount of time (10$^4$ yr). Here, we find only 1 confirmed DE-RSG, with the other objects in the original vL05 sample having normal RSG mass-loss rates. Searching the \citet{davies2018humphreys} catalogue of CSGs in the LMC yields one other candidate, LI-LMC 1100. If these two objects do represent a specific phase in stellar evolution, during which RSGs undergo extreme levels of mass-loss, it must be extremely short lived ($\sim$10$^{4}$ yr\footnote{WOH G64 and LI-LMC 1100 represent only 3\% of the RSG population in the LMC, and if we assume the RSG phase is 10$^{6}$ yrs \citep[e.g.]{georgy2012yellow} then the DE-RSG phase would last on the order of 10$^{4}$ yrs.} ) and would only be able to peel away an additional $\leq$1\msun\ of material.\footnote{From this study, it is not clear whether these DE-RSGs do represent a real, short-lived phase for single stars, or whether they are the product of unusual evolutionary paths such as post-mergers, mass-transfer systems, or common envelopes.} This effectively rules out the single-star scenario for the production of post-RSG stripped stars and Type Ibc SN. 

\begin{figure*}
    \centering
    \includegraphics[width=15cm]{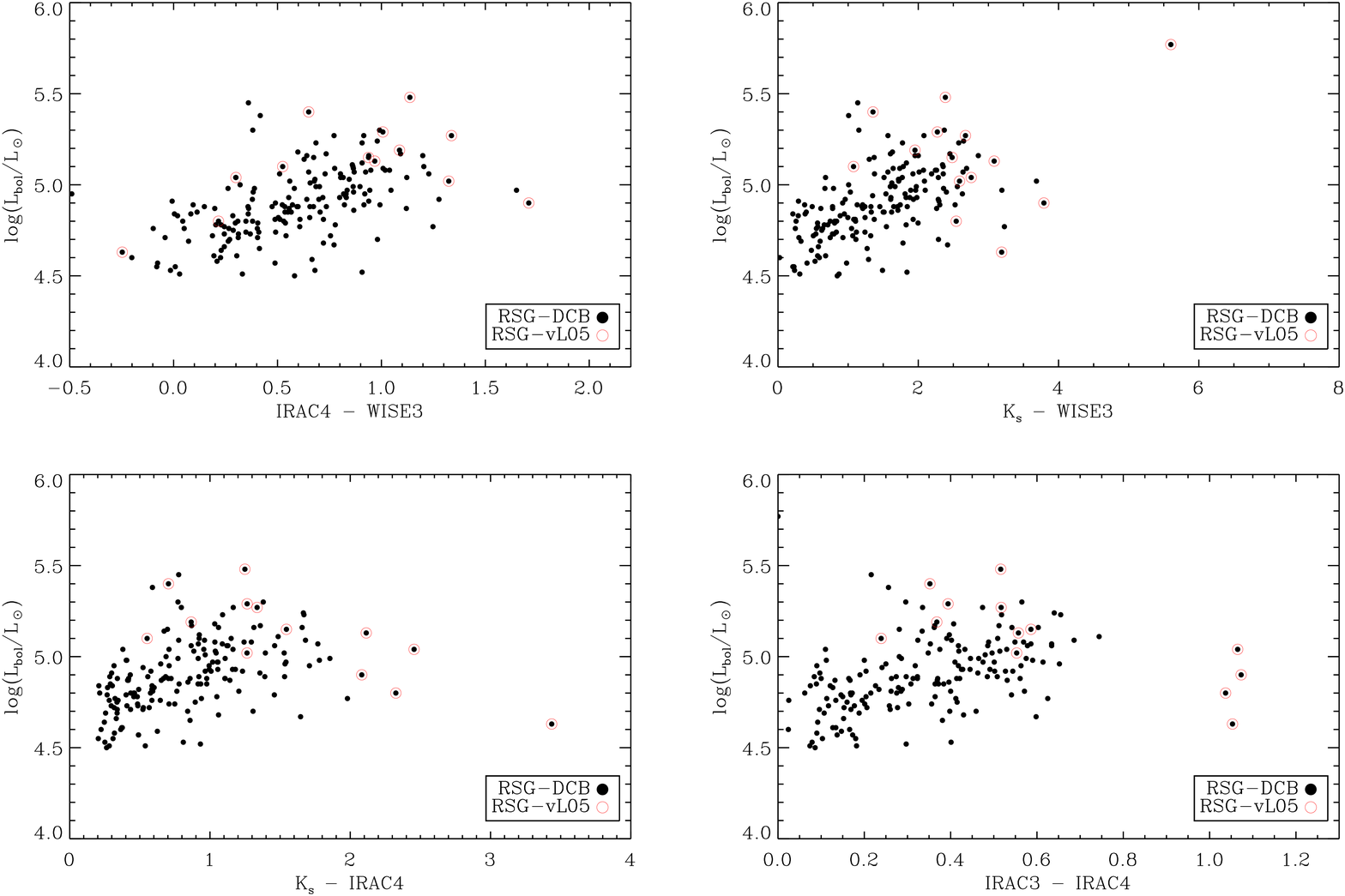}
    \caption{Colour luminosity diagrams used  comparing the distribution of the vL05 DE-RSG sample to all M-type supergiants in the DCB18 catalogue. The red outlines show the objects originally included in the vL05 sample, while the black dots show the objects in the DCB18 LMC catalogue.}
    \label{fig:cc_all}
\end{figure*}

\begin{figure*}
    \centering
    \includegraphics[width=15cm]{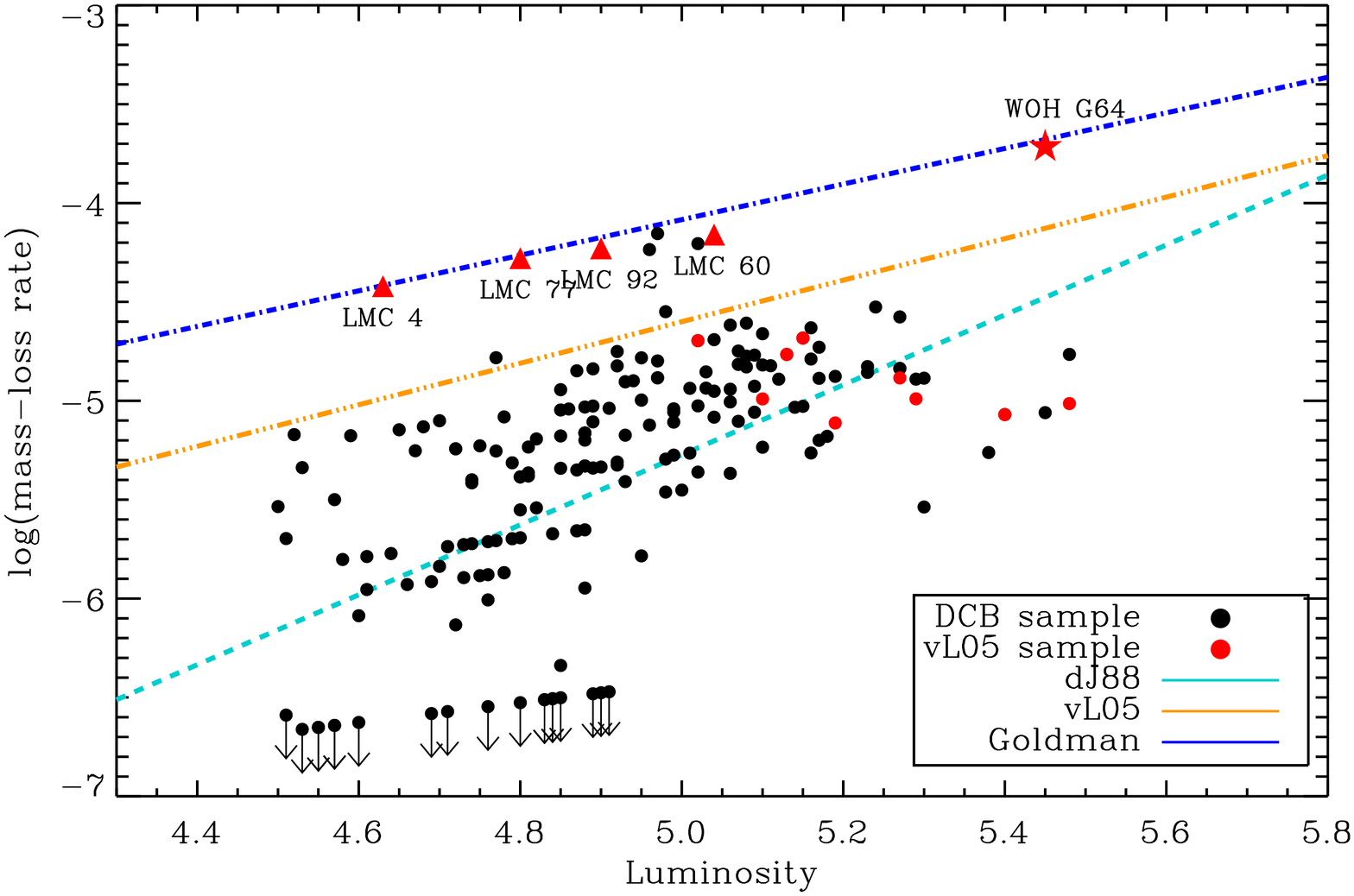}
    \caption{Mass-loss rates and luminosities for the entire LMC M-type star population. Here, the red symbols are the objects included in the vL05 prescription, and the black symbols are the entire LMC population. We highlight the likely AGB stars with red triangles, and the bona-fide DE-RSG WOH G64 with the star. We also overplot some \mdot-prescriptions commonly used in evolutionary models. With the exception of WOH G64, no object in the LMC falls into the category of DE-RSG as defined in Section \ref{sec:define}. }
    \label{fig:lmc_mdots}
\end{figure*}

In single-star evolutionary models of roughly 20 - 30 $M_{\odot}$ stars, any post-RSG phase as an LBV or a WR is likely an artifact of using the unrealistically high \mdot-prescription from vL05 \citep[e.g.][]{ekstrom2012grids,chieffi2013pre} for a sustained period of time. In addition, when comparing the DE-RSGs to all M-type supergiants in the LMC, we show in Fig. \ref{fig:cc_all} that the DE-RSGs from vL05 do not stand out as a distinct group of stars in any colour-magnitude space, thus making it unlikely that they represent a specific phase of single stellar evolution\footnote{While the IRAC3 - IRAC4 colour does show 4 objects as being significantly reddened compared to the general population, these are the 4 suspected AGBs. When considering only stars with luminosities about \logl=5.0, i.e. those which are definitely RSGs, there is no clear distinction between the vL05 sample and the M-type population as a whole.}. We also calculate \mdot\ values for the entire LMC M-type star population and again, the vL05 sample do not clearly stand out as a unique group of objects in this space, see Fig. \ref{fig:lmc_mdots}. We also plot the vL05 mass-loss rate prescription for DE-RGs in Fig.~\ref{fig:lmc_mdots}, where it can be seen that this prescription is not representative of either the DE-RSGs or the RSGs as a whole.  Instead, the vL05 prescription is significantly above most of the proposed DE-RSGs, representing an upper threshold to the main group, exceeded only by the four AGB stars and WOH G64.

Given this, we suggest that the vL05 prescription or other recipes for ``enhanced'' RSG mass loss should not be employed in single-star evolutionary codes. Single star codes with such high mass loss will produce erroneous outcomes. Without invoking this high mass-loss, RSGs in single-star models will not be able to produce post-RSG LBVs, BSGs, and WR stars. With single-star channels unable to make stripped stars via wind mass loss, we suggest that all $M_{\rm ZAMS} \la 40 M_{\odot}$ progenitors to stripped-envelope SNe (Types Ibc and IIb) come from binary evolution. This is consistent with several lines of evidence from studies of SE-SNe \citep{drout11,smith11,hachinger12,smith2014mass,lyman16}.  Whether or not initially more massive single stars can yield WR stars via wind mass loss is a separate topic that is not constrained by our work here, since those stars do not evolve through a RSG phase. %this is a pre-emptive comment for jorick, so that he doesn't lash out at us in self defence

\section{Conclusions}
In this paper we have analyzed the SEDs and re-appraised the mass-loss rates for the proposed sample of DE-RSGs from \citet{van2005empirical} using the updated LMC CSG catalogue of \citet{davies2018humphreys}. Specifically, we examined the 14 RSGs with luminosity above 10$^{4.6}$ L$_{\odot}$ in this sample, all of which are also included in the DCB catalogue. Using {\tt DUSTY} modelling, we find that only one (WOH G64) of the original 14 RSGs in the vL05 sample actually has substantial mid-IR excess or circumstellar extinction. From our work, the main conclusions and implications are:
\begin{itemize}
    \item We identify only one RSG in the LMC that can be confirmed as truly dust enshrouded (WOH G64),  with a mass-loss rate exceeding 10$^{-4}$ $M_{\odot}$ yr$^{-1}$.  We also identify one other DE-RSG candidate (LI-LMC 1100) that was not included in the original sample of proposed DE-RSGs. The nature of LI-LMC 1100 cannot currently be confirmed without further observation, but from a color-luminosity plot, the object can be identified as being luminous and extremely red. It's position within a 7 -- 10 Myr old cluster suggests that it is not an embedded protostar.
    \item The rest of the objects in the original vL05 sample cannot fairly be described as DE-RSGs, and instead appear to be normal RSGs with little mid-IR excess and relatively low mass-loss rates.
    \item Including both of the above objects, we classify only 3\% of the RSG population in the LMC as dust enshrouded. This severely restricts the duration and evolutionary role of dust-enshrouded phases. If this is an evolutionary phase undergone by all RSGs, we find that it must be extremely short lived ($<$30,000 yrs) and would only be able to remove up to 1--2\msun\ of material. This is not enough to drive a single star back to the blue of the HRD. 
    \item Overall, this means that low-luminosity WR stars, blue supergiants, and  LBVs are unlikely to arise from post-RSG evolutionary phases in single stars. It also means that the relatively low initial  mass progenitors of Type Ibc and Type IIb SNe ($M_{\rm ZAMS} \la 40 M_{\odot}$) cannot form via a single-star pathway.
    
    \item We suggest that the prescription of vL05 does not describe a truly persistent high mass-loss phase experienced by RSGs, and hence {\it should not} be implemented in single-star evolutionary models, unless one wishes to use models to investigate effects of a hypothetical, short-lived burst of mass loss prior to SN.
\end{itemize}

%Here, we have compared the projected separations of two groups of RSGs in the LMC to the nearest O-star, following the method of \citep{smith16}. The first group was a complete sample of RSGs taken from \citet{davies2018humphreys}, while the second was the highly dust enshrouded (and high mass-loss) RSGs from the sample of \citet{van2005empirical}. We find that the dust enshrouded RSGs from the \citeauthor{van2005empirical} sample have similar age indicators to the lower mass RSGs in the LMC, but have a luminosity distribution skewed to the highest \lbol\ values. We suggest two explanations for our findings,
%\begin{itemize}
  %  \item The RSGs in the van Loon et al. sample represent the most evolved, low mass RSGs, currently undergoing an episode of extreme mass-loss prior to SN. 
 %   \item Alternatively, the objects in the RSG-vL sample could be descended from binary interactions, such as mass gainer or merger events, explaining their larger separations from the general RSG population in the LMC. The rejuvination from these interactions could explain the skewing of the \lbol-distribution to higher luminosities.
%\end{itemize}

%In either case, we have demonstrated clearly that the high mass-loss rates of \citet{van2005empirical} are not applicable to the majority of the RSG phase, and thus should not be implemented as normal RSG \mdot\ in stellar models. 
\section*{Acknowledgements}

The authors thank the referee for insightful discussion which helped improve the manuscript. The authors would also like to thank Ben Davies for useful comments and discussion. Support for this work was provided by NASA through Hubble Fellowship grant HST-HF2-51428 awarded by the Space Telescope Science Institute, which is operated by the Association of Universities for Research in Astronomy, Inc.,
for NASA, under contract NAS5-26555. 

%The Acknowledgements section is not numbered. Here you can thank helpful colleagues, acknowledge funding agencies, telescopes and facilities used etc. Try to keep it short.

\appendix
Here we present the results of dust shell modelling for all 14 RSGs included in this study, Figs. 7 \& 8. In these figures, the black crosses represent the observed photometry, synthetic photometry is shown by the yellow circles, the solid green line shows the best {\tt DUSTY} model fit and the pink dashed line shows the flux from dust emission only. We also show Spitzer-IRS data where available, indicated by the solid red line, which is generally well matched by the best-fit models and photometry. In all cases we see that the mid-IR excess is weak, implying lower mass-loss rates than found by \citet{van2005empirical}. For LI-LMC 4, LI-LMC 60, LI-LMC 70 and LI-LMC 92 we were not able to fit both the short and long wavelength flux, probably because these stars are AGB stars with lower photospheric temperatures than the RSG temperatures used in our model grid, but it can be seen from the 10$\mu$m emission that the dust shell mass is low. 

\begin{figure*}
    \centering
    \includegraphics[width=8cm]{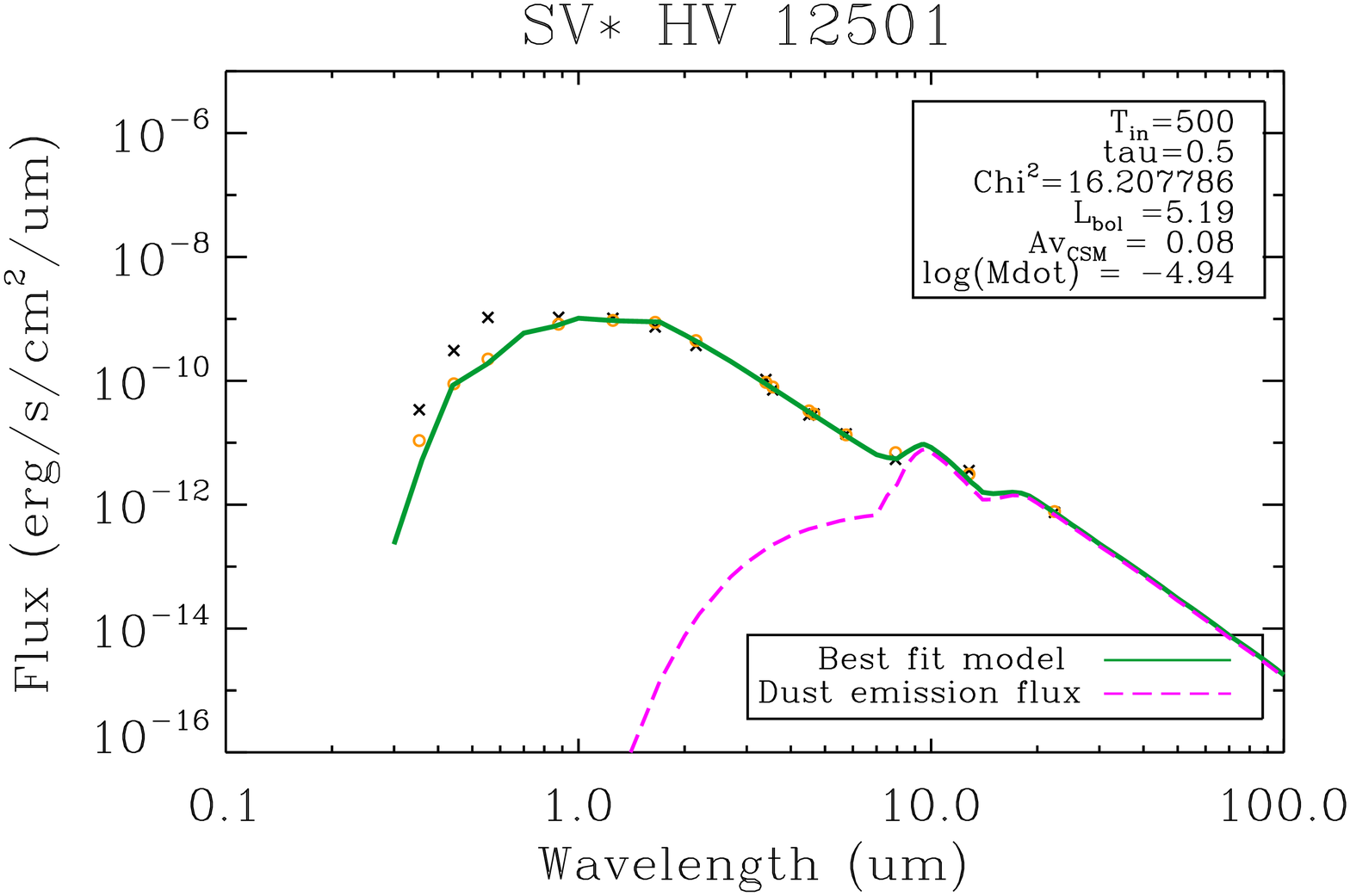}\includegraphics[width=8cm]{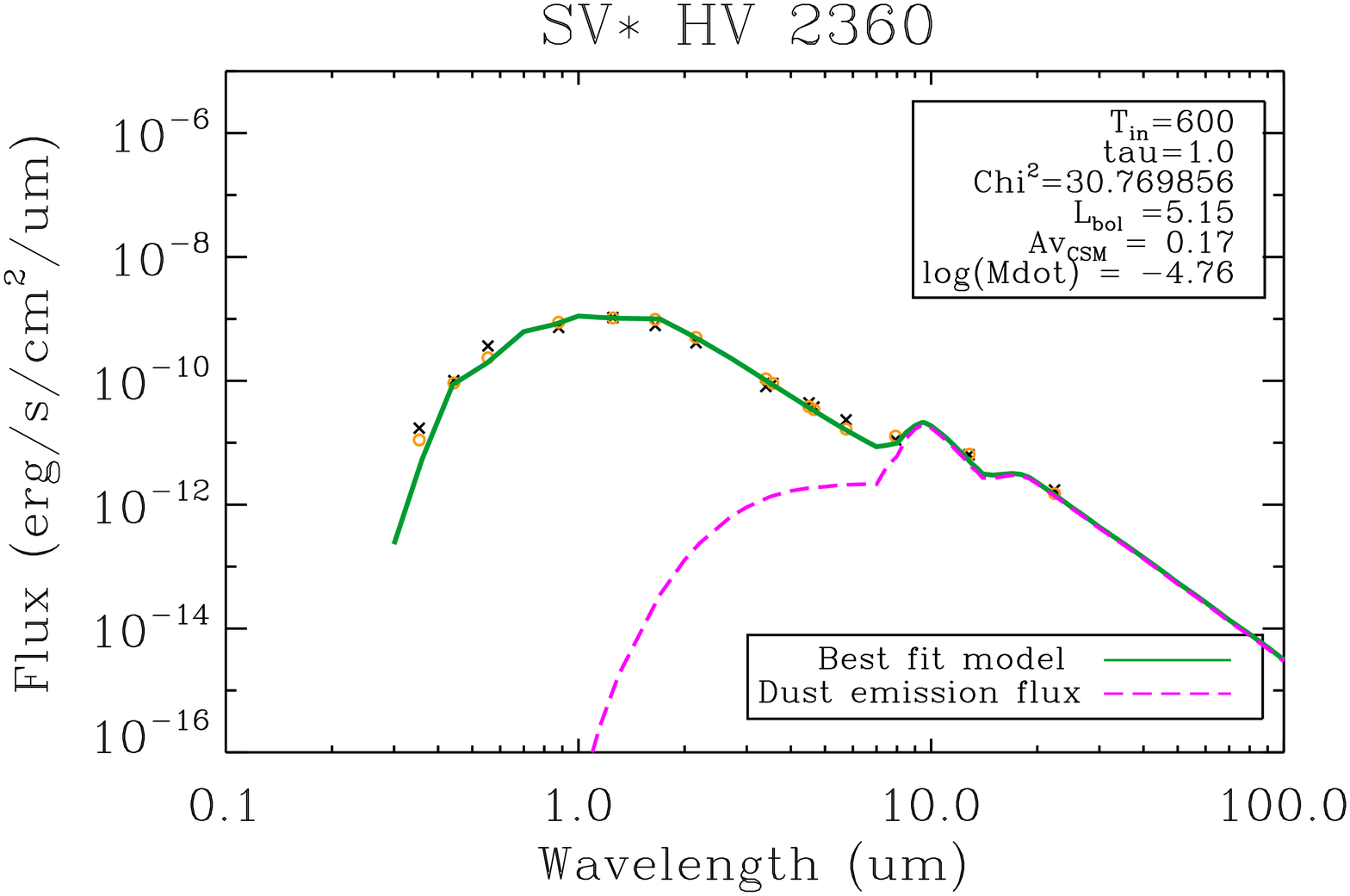}
    
    \includegraphics[width=8cm]{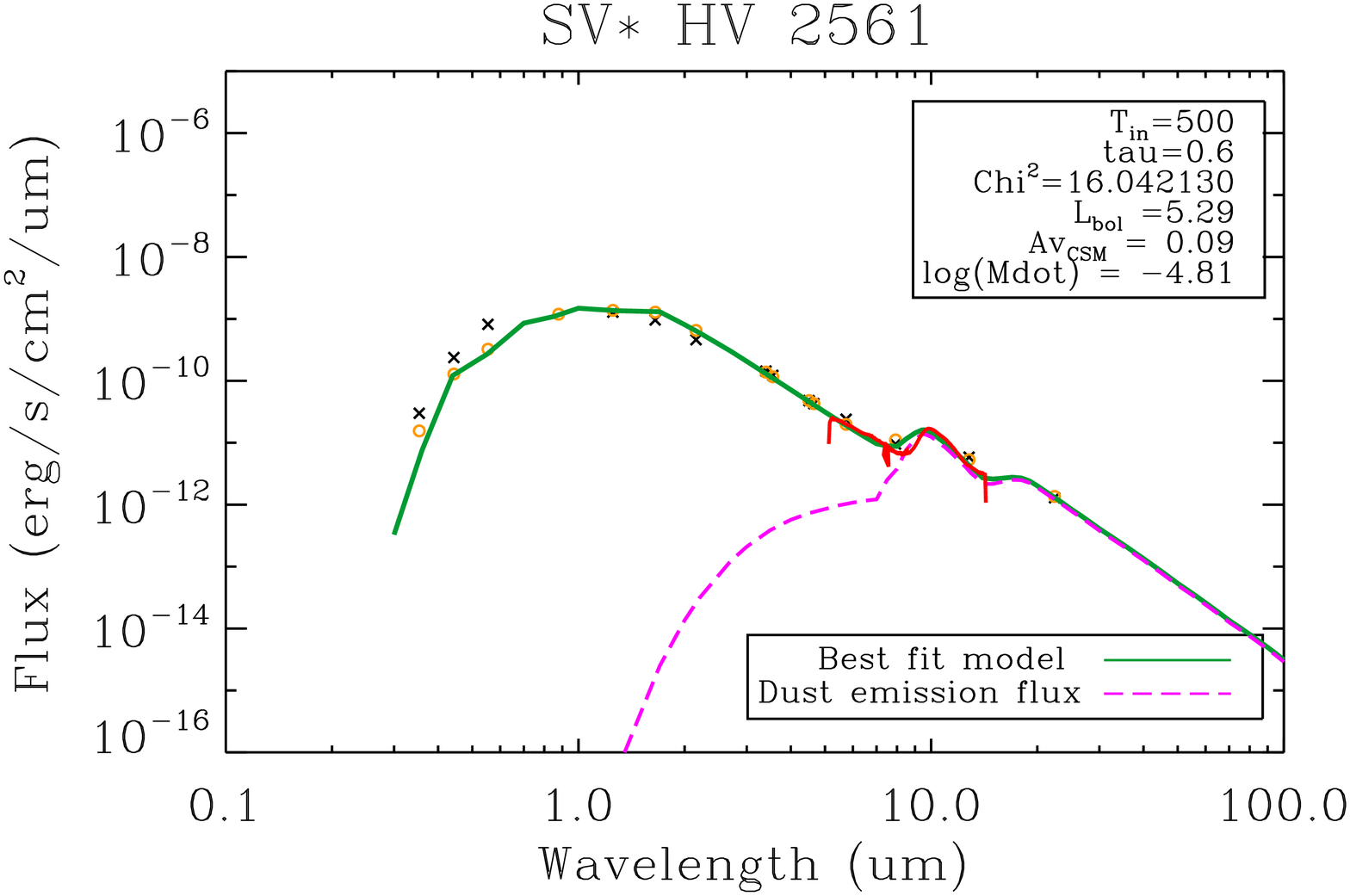}\includegraphics[width=8cm]{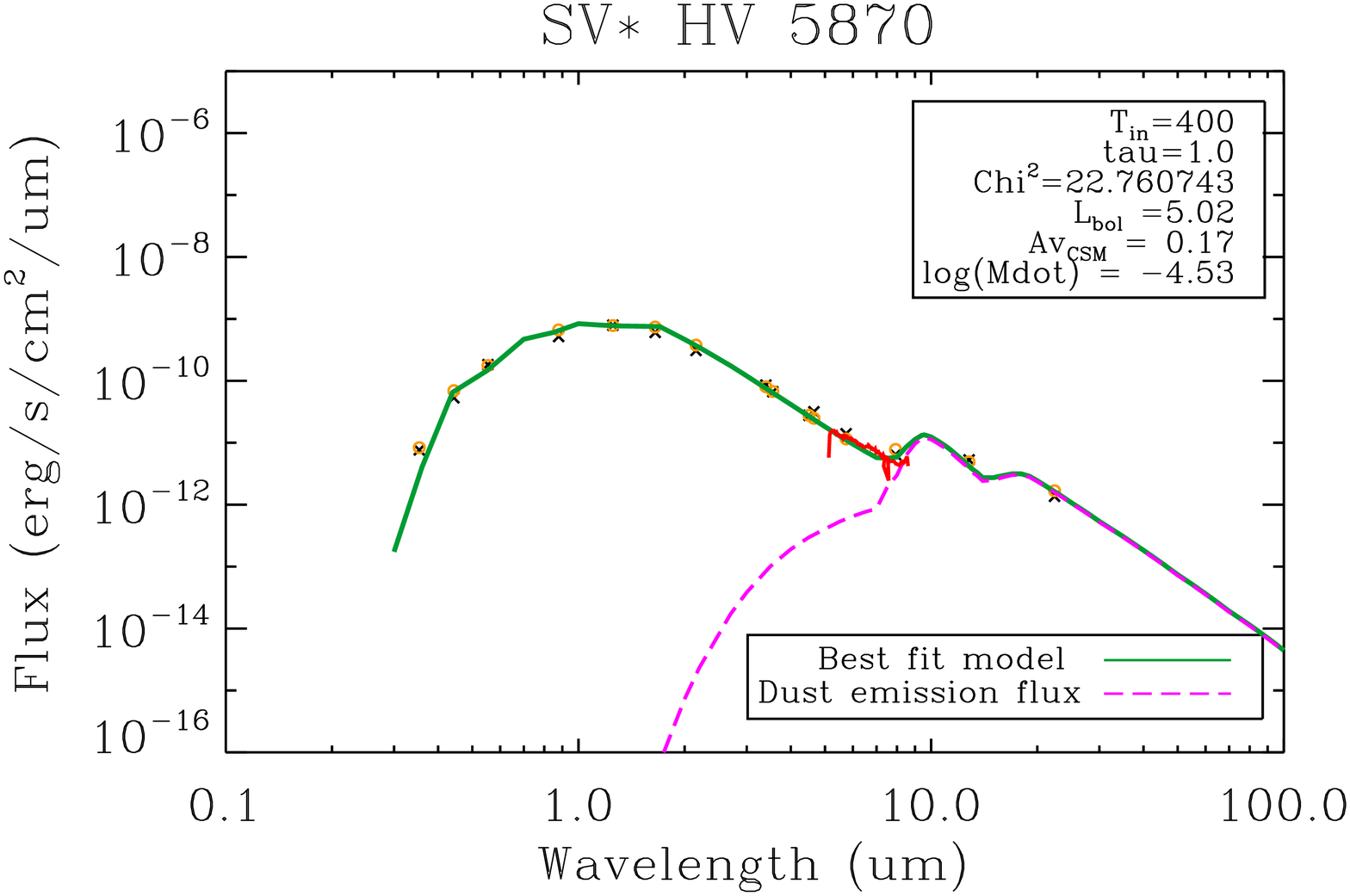}
   
    \includegraphics[width=8cm]{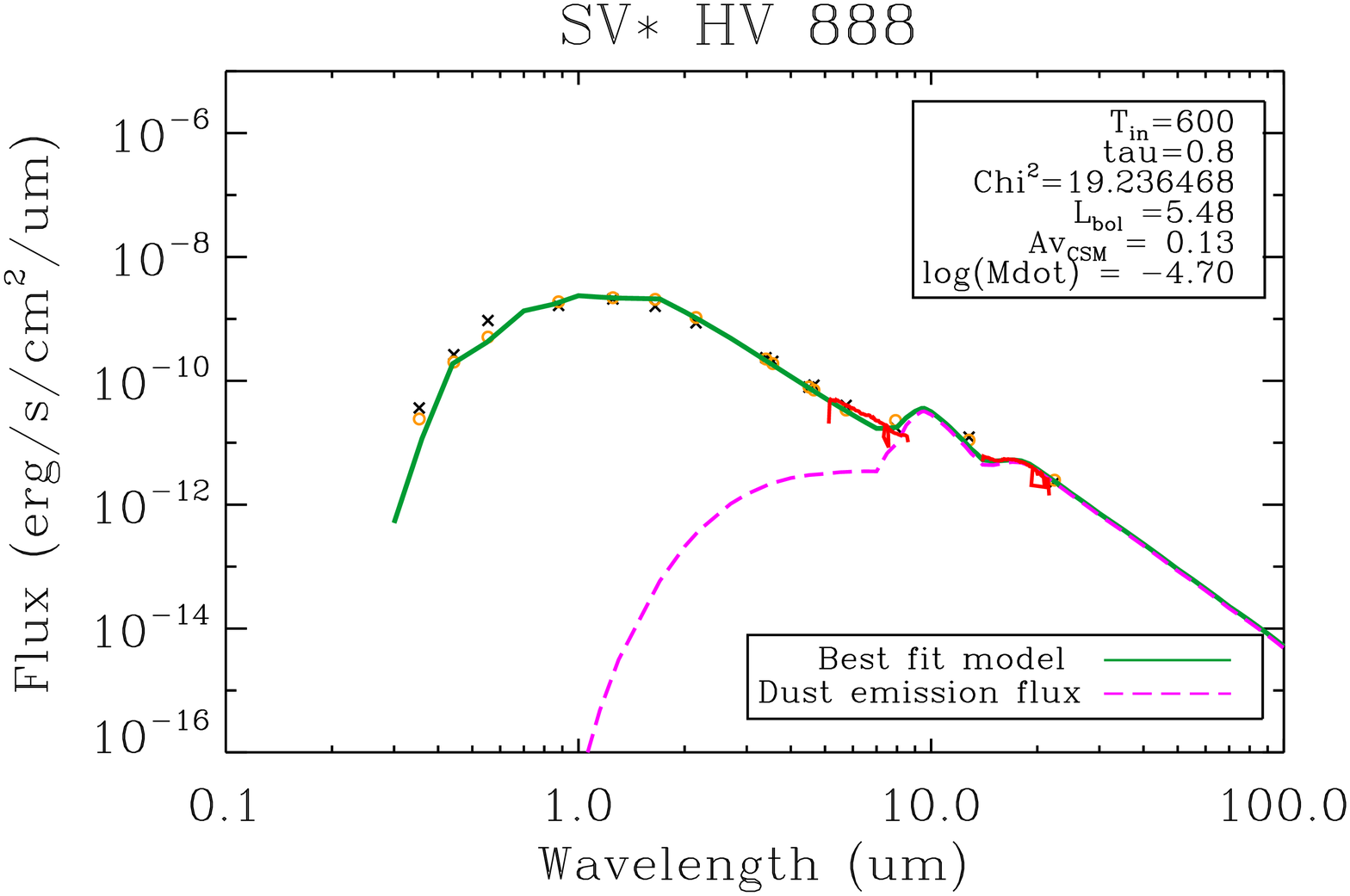}\includegraphics[width=8cm]{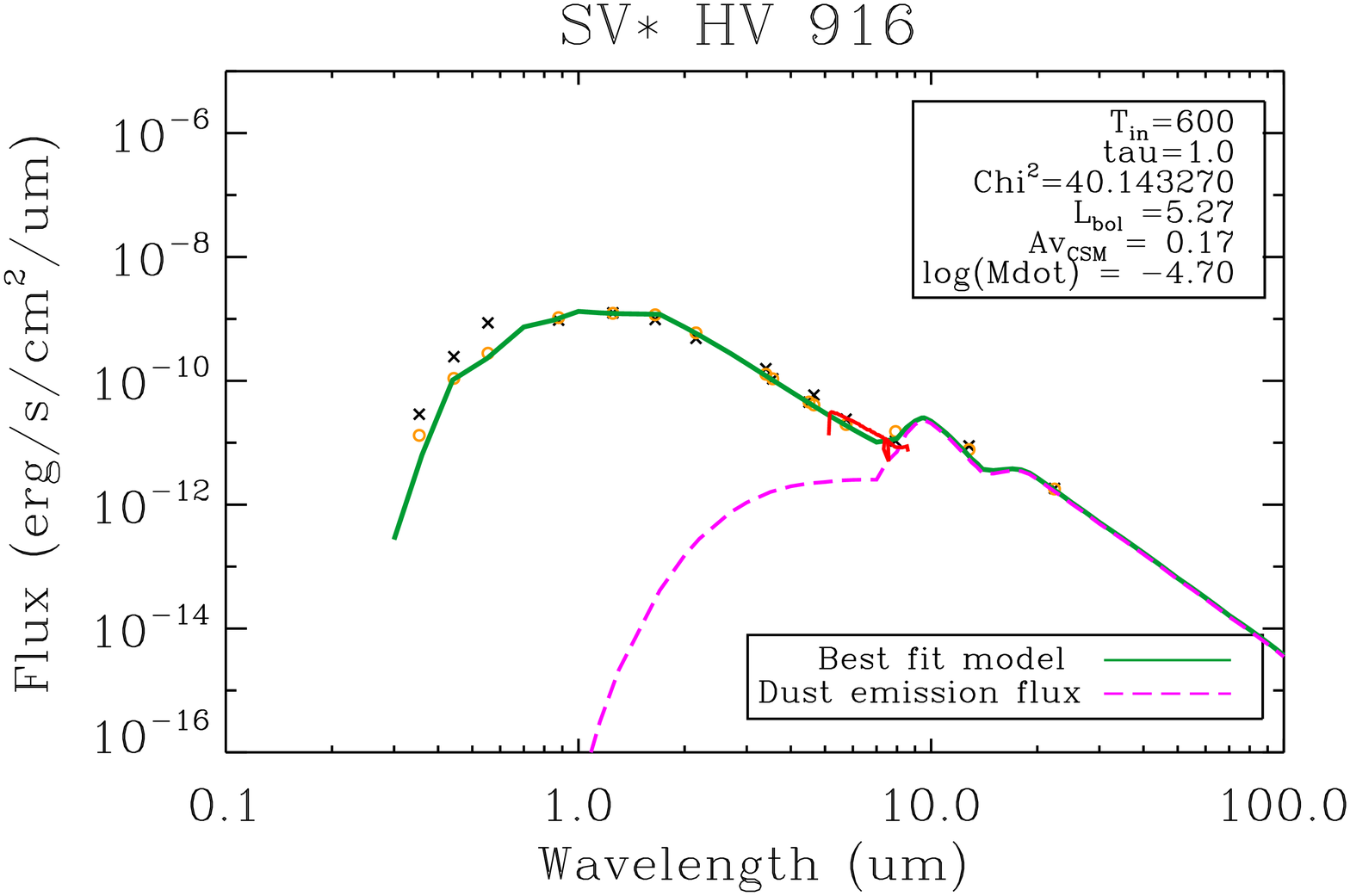}
   
    \includegraphics[width=8cm]{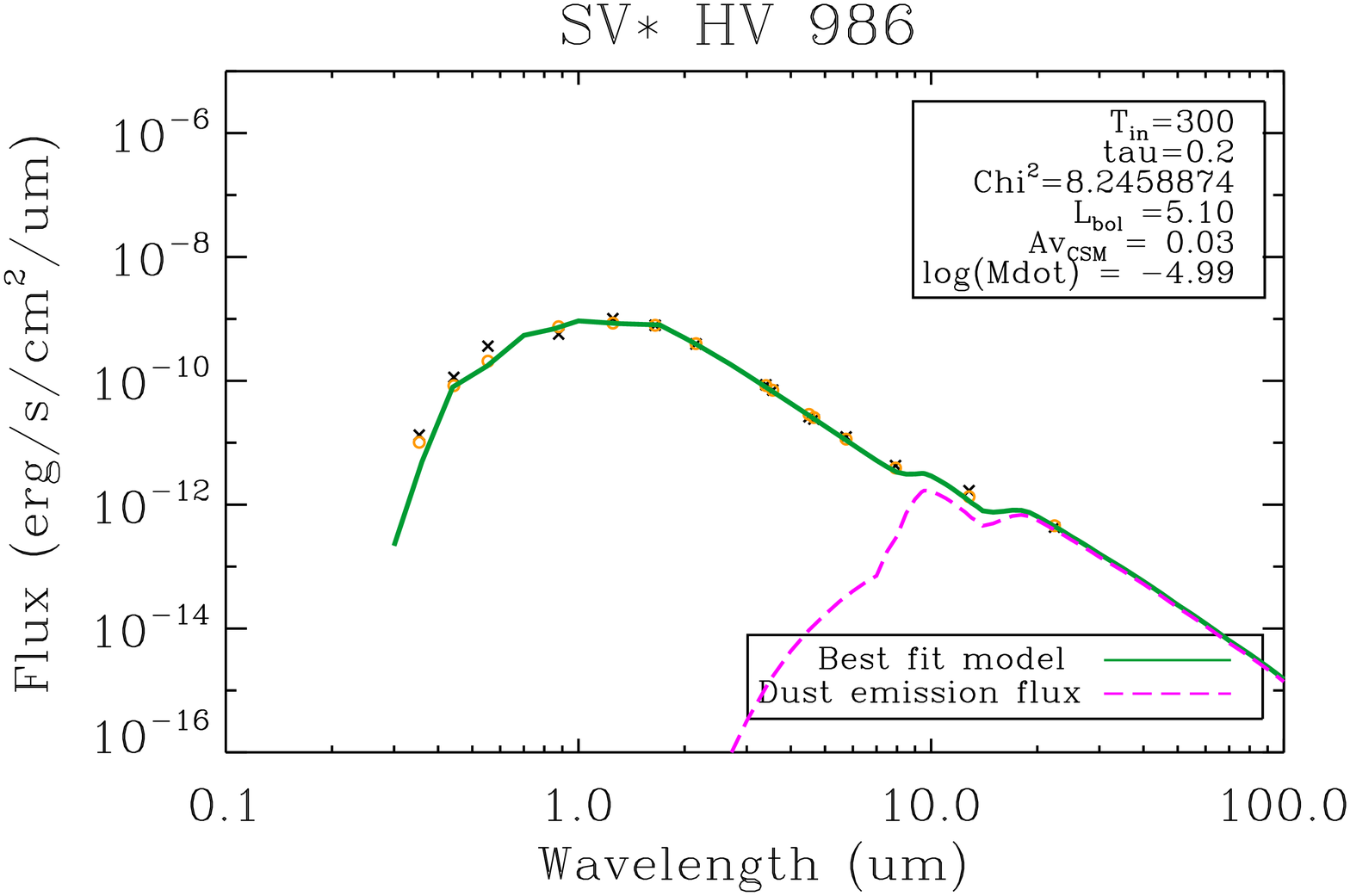}\includegraphics[width=8cm]{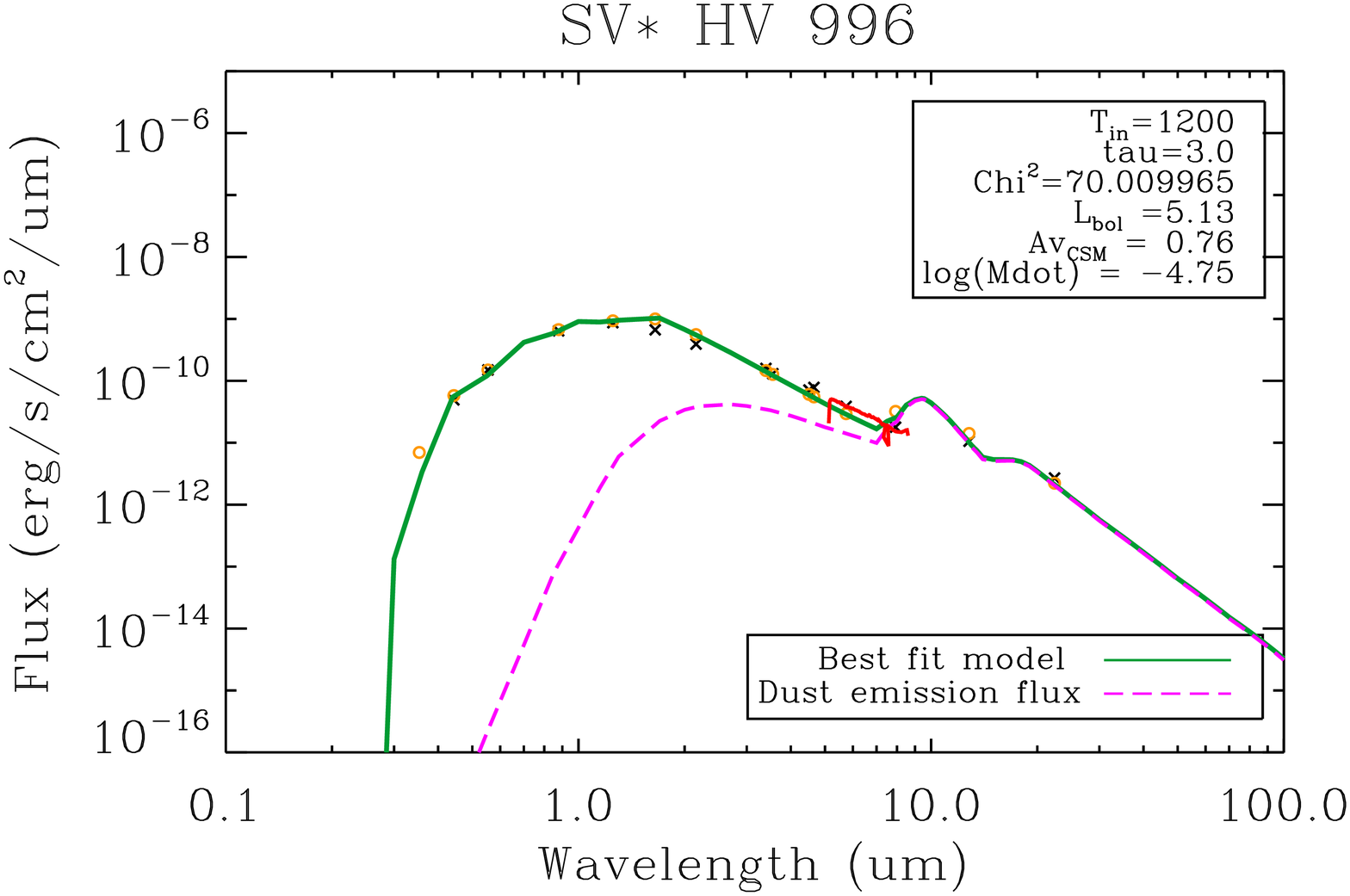}

    \caption{The  best  fit  model  for each of the putative DE-RSGs. The  solid green line shows the best model fit from this work and the pink dashed line shows the flux from dust emission. Where available, we also show Spitzer-IRS data (solid red line) though we do not use this for analysis.}
   \label{fig:app1}
\end{figure*}
\begin{figure*}
    \centering
    
        \includegraphics[width=8cm]{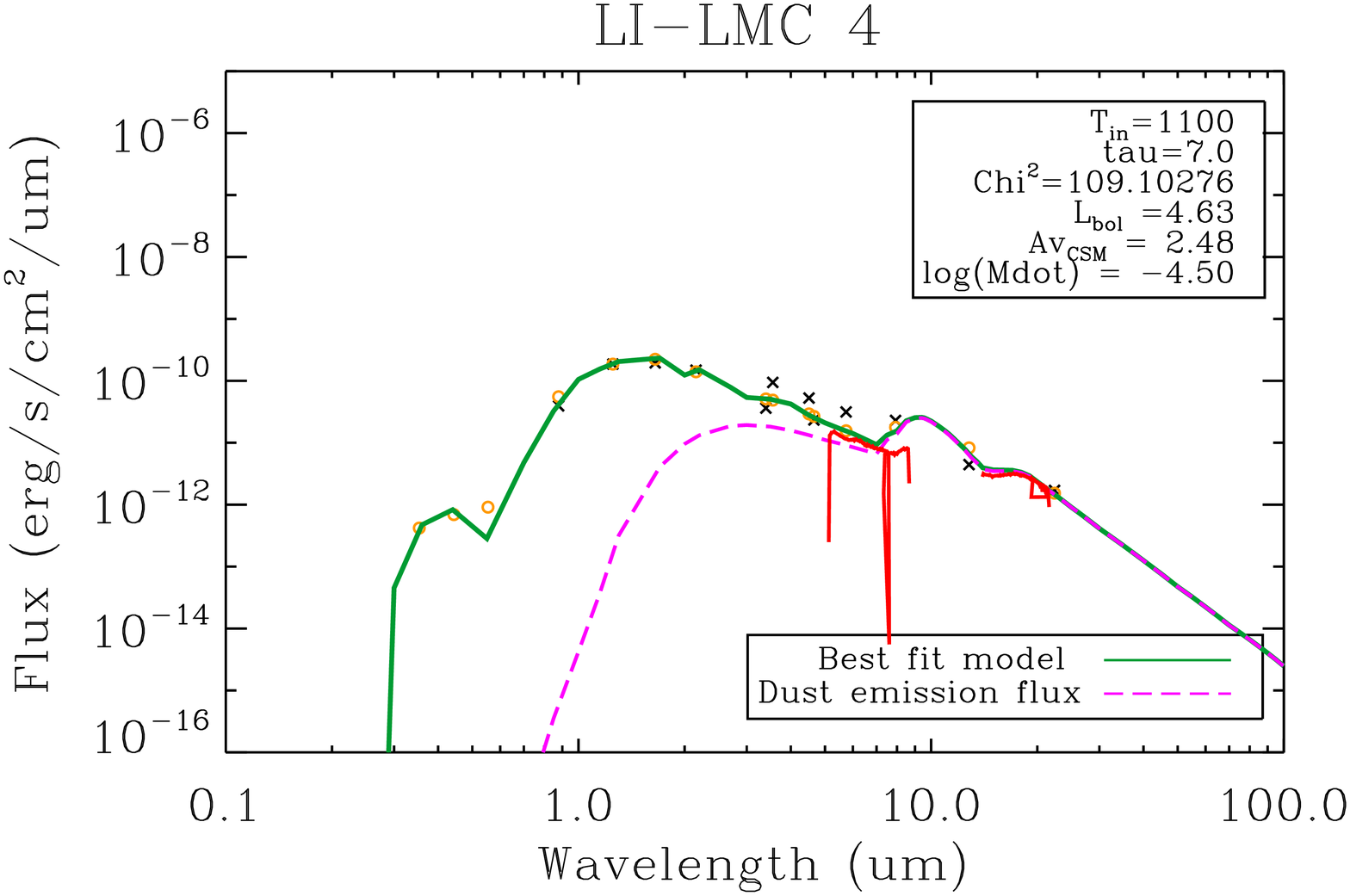}\includegraphics[width=8cm]{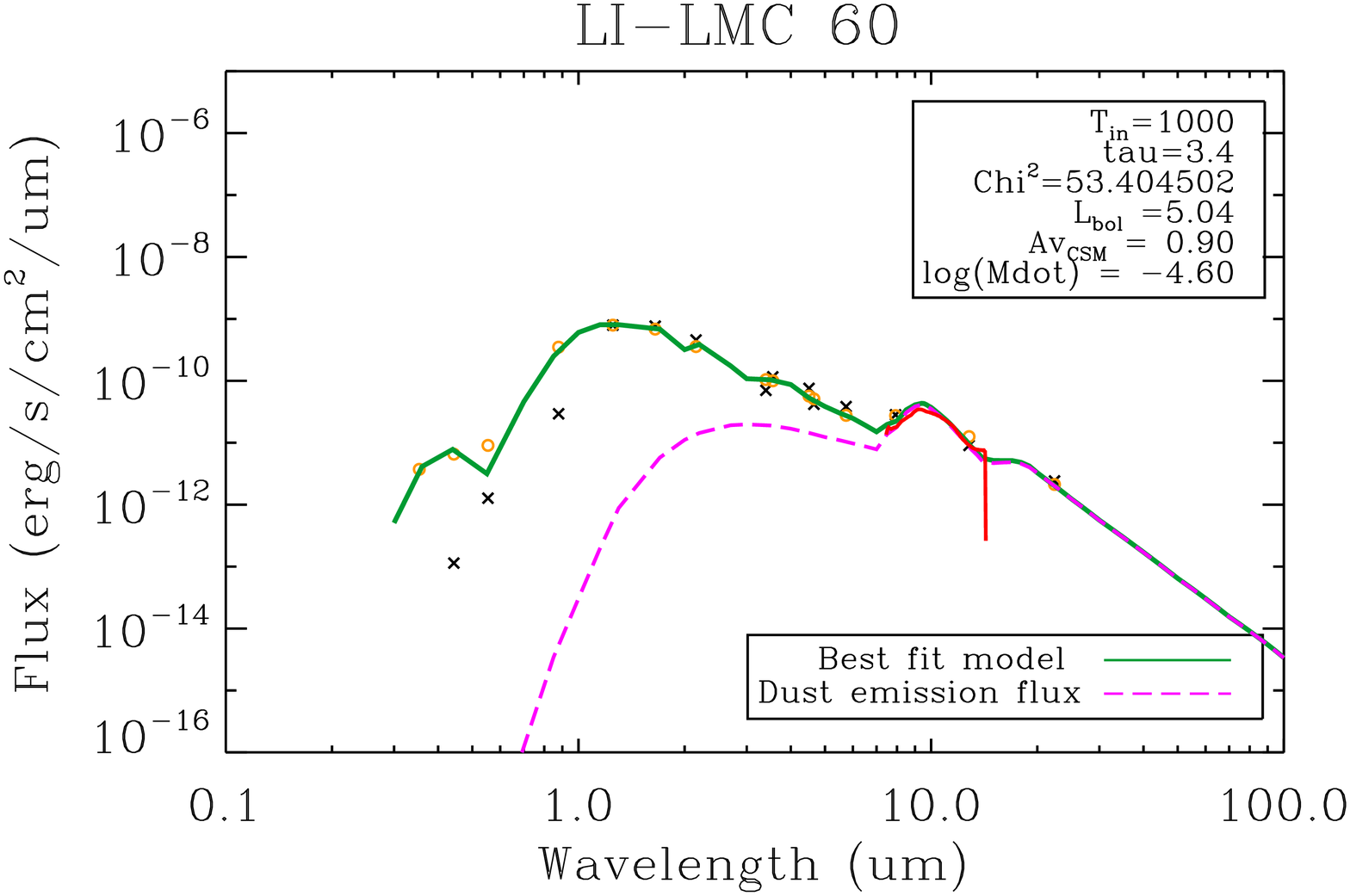}
   \includegraphics[width=8cm]{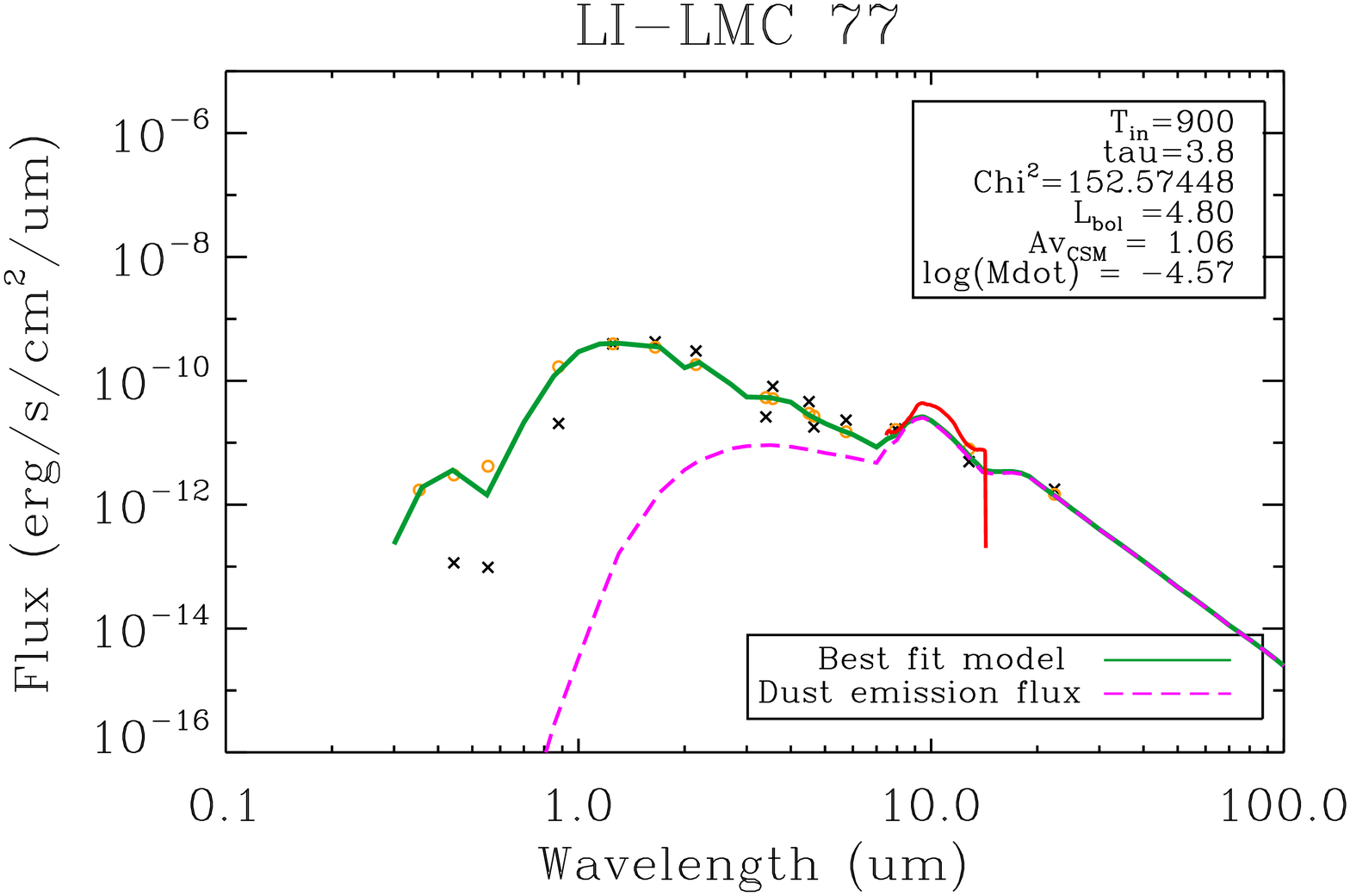}\includegraphics[width=8cm]{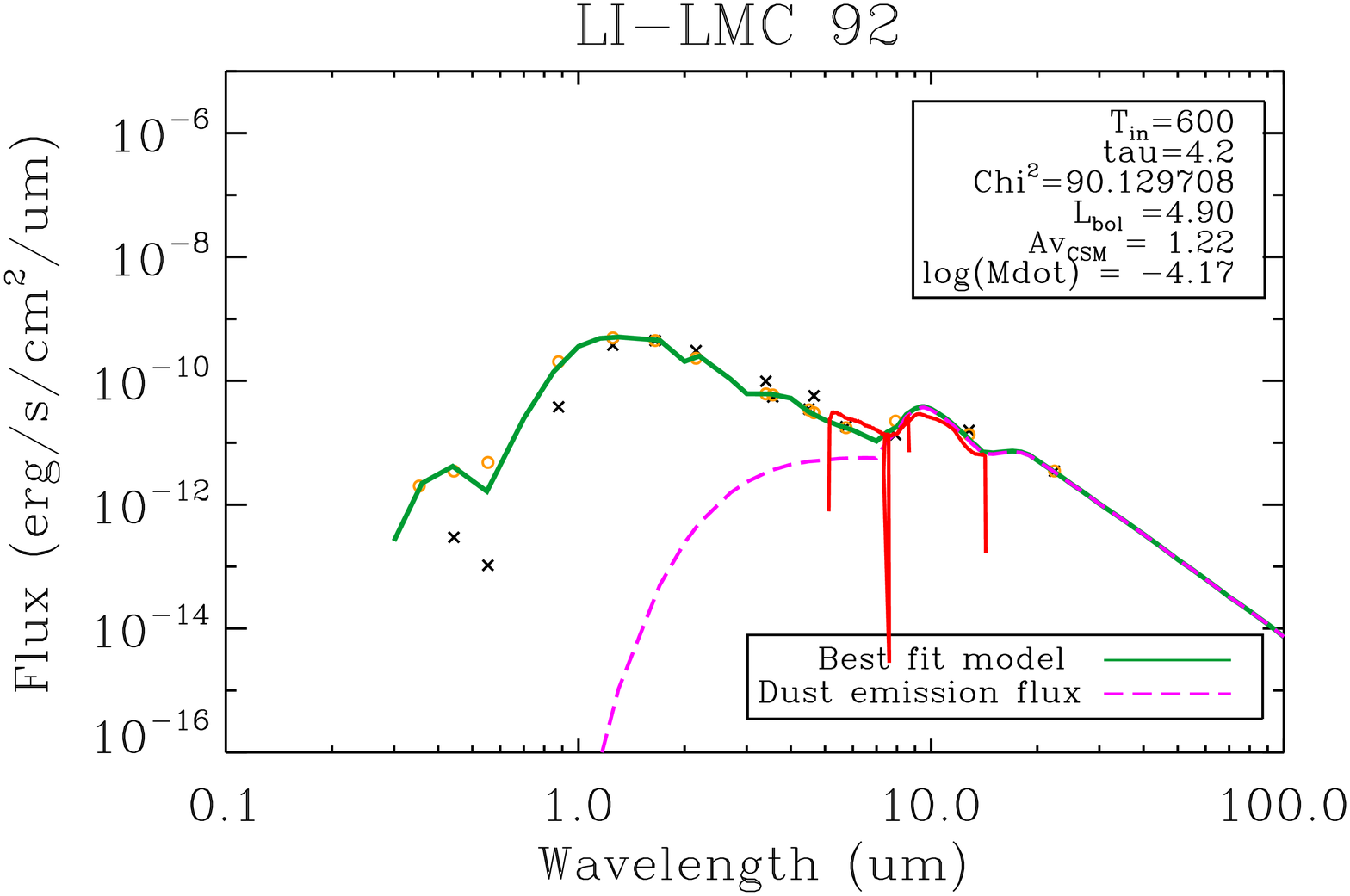}
  
    \includegraphics[width=8cm]{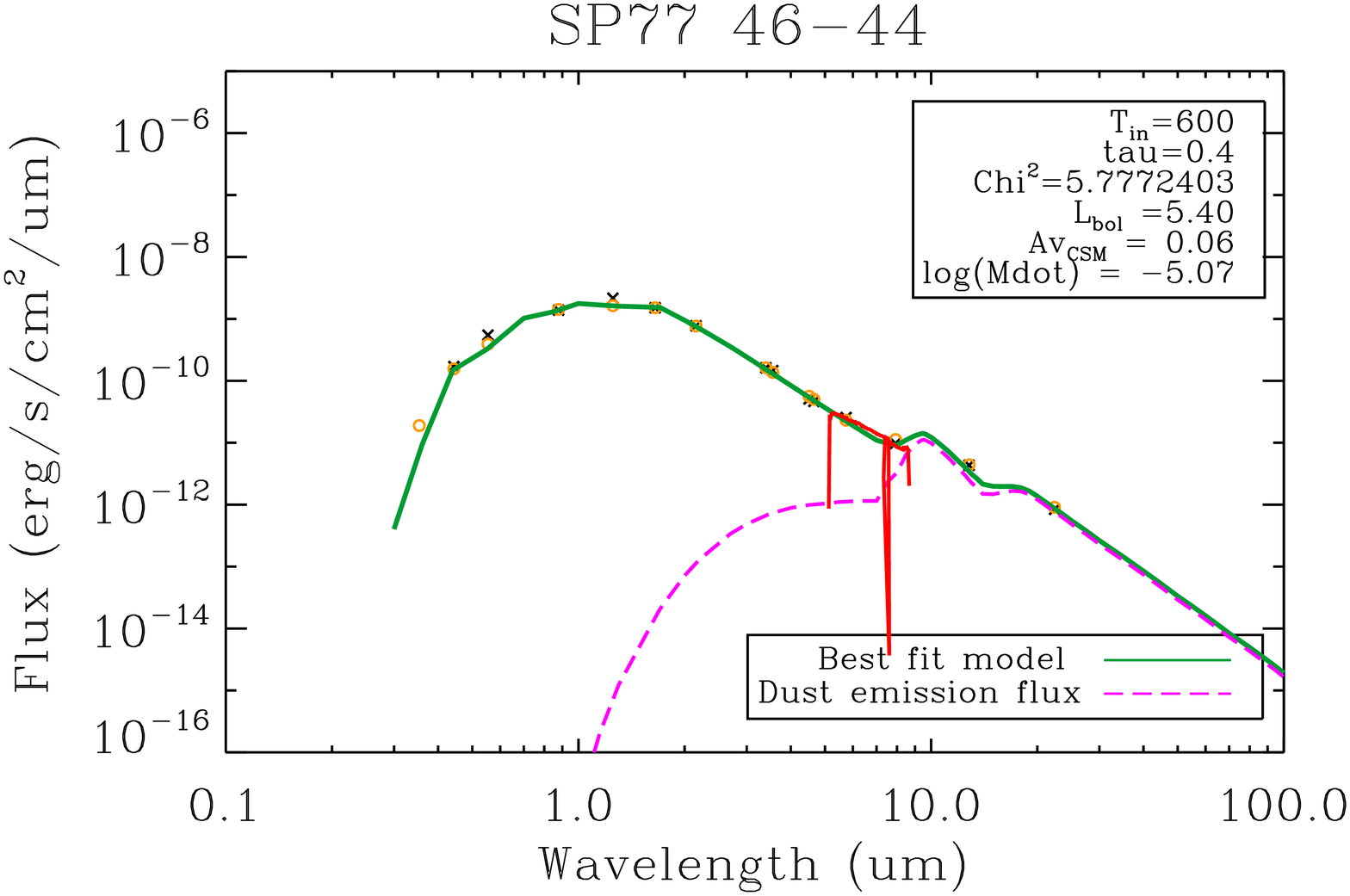}\includegraphics[width=8cm]{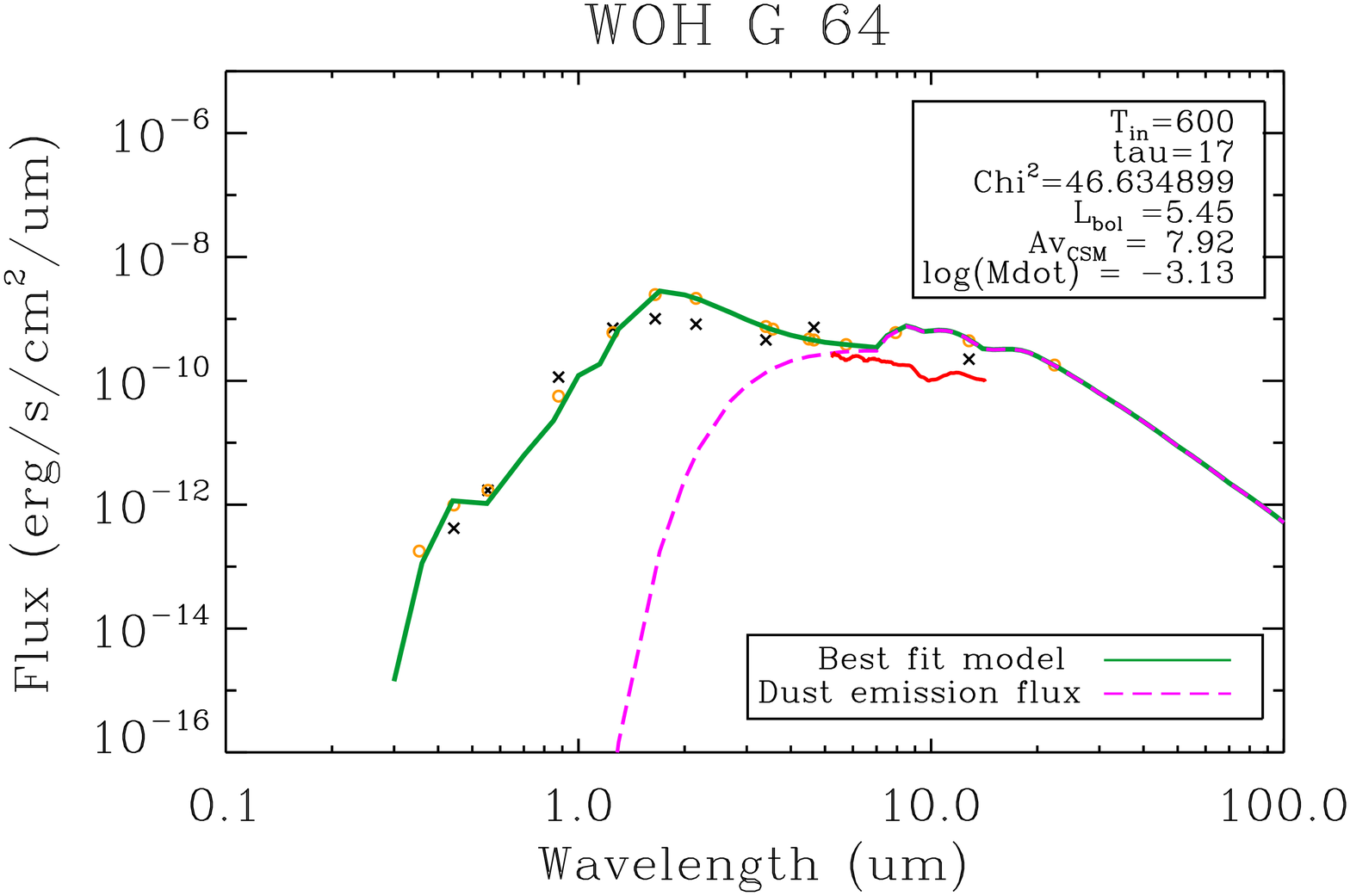}
    \label{fig:appp2}
    \caption{Same as previous figure.}

\end{figure*}


\bibliography{sample63}{}
\bibliographystyle{aasjournal}



\end{document}